\documentstyle[12pt,psfig]{article}
\begin{document}

\vskip 1truecm
\rightline{Preprint  MCGILL-98/7}
\rightline{ e-Print Archive: hep-ph/9805264}
\vspace{0.2in}
\centerline{\Large Measuring the Broken Phase }
\vspace{0.15in}
\centerline{\Large Sphaleron Rate Nonperturbatively} 
\vspace{0.3in}

\centerline{\large Guy D. 
	Moore\footnote{e-mail: guymoore@physics.mcgill.ca}}

\medskip

\centerline{\it Dept. of Physics, McGill University}
\centerline{\it 3600 University St.}
\centerline{\it Montreal, PQ H3A 2T8 Canada}

\medskip

\centerline{\bf Abstract}

We present details for a method to compute the broken phase sphaleron
rate (rate of hot baryon number violation below the electroweak phase
transition) nonperturbatively, using a combination of multicanonical and 
real time lattice techniques.  The calculation includes the ``dynamical
prefactor,'' which accounts for prompt 
recrossings of the sphaleron barrier.  The prefactor
depends on the hard thermal loops, getting smaller with increasing Debye
mass; but for realistic Debye masses the effect is not large.  The
baryon number erasure rate in the broken phase is slower than a
perturbative estimate by about $\exp(-3.6)$.  Assuming the electroweak
phase transition has enough latent heat to reheat the universe to the
equilibrium temperature, baryon number is preserved after the phase
transition if the ratio of (``dimensionally reduced'' thermal)
scalar to gauge couplings $\lambda / g^2$ is less than .037.

\section{Introduction}

20 years ago, t'Hooft showed that baryon number is not a good quantum
number in the standard model \cite{tHooft}.  The reason involves the
nontrivial vacuum structure of the SU(2) (weak) gauge group of the
standard model.  In any gauge theory, the vacuum is not unique; any
gauge transformation of $\vec{A} = 0$ has zero energy and is an
acceptable vacuum.  But SU(2) (and any simple gauge group) has the
property that the space of 3-D gauge transformations is topologically
nontrivial.  A gauge transformation has an integer $\pi_3$ winding
number associated with it.  Since the winding number must be an integer,
the space of smooth gauges, and also the space of vacua, is
disconnected.  The different connected components are characterized by
their values of Chern-Simons number,
\begin{equation}
N_{\rm CS} \equiv \frac{ g^2 }{32 \pi^2} \int d^3 x \epsilon_{ijk} 
	 \left( F_{ij}^a A_k^a - \frac{g}{3} f_{abc} A^a_i A^b_j A^c_k
	\right) \, ,
\end{equation}
which is an integer for a vacuum configuration, though not necessarily
for an excited state.  Classically, for the gauge fields to change from
one topological vacuum at time $t = t_i$ to another at time $t=t_f$,
they must pass through excited states in the intervening time; to be
specific, 
\begin{equation}
dN_{\rm CS}/dt = \frac{g^2}{32 \pi^2} \int d^3 x F_{\mu \nu}^a 
	F^{\mu \nu}_a \, ,
\end{equation}
which is clearly a gauge invariant quantity (though its integral,
$N_{\rm CS}$, is not, because of the constant of integration).
This means that it is possible to pass from a vacuum configuration to a
gauge copy of that configuration via a path which cannot be smoothly
deformed to remain always in vacuum.  If we mod out the space of 3-D
configurations by the gauge transformations, the space we get will then
have noncontractible loops.

This topological structure appears both in SU(2) (weak) 
and SU(3) (strong), where it
is responsible for the physics of spontaneous chiral symmetry breaking.
What t'Hooft noticed is that, because fermions couple to the weak 
SU(2) group of the standard model chirally,
the anomaly relates $N_{\rm CS}$ to baryon number.  If the gauge fields
pass through some nonvacuum intermediate state from one topological 
vacuum to another (or around a noncontractible loop, if we think of
configurations modulo gauge transformations), 
baryon number changes.  Such changes are classically
forbidden at zero temperature, so they only occur via quantum tunneling.
Because the SU(2) gauge coupling is weak, and because the Higgs field
breaks the symmetry, such processes are steeply exponentially
suppressed, by $\sim \exp(- 16 \pi^2 / g^2) \sim 10^{-170}$.  Hence such
processes are of no terrestrial phenomenological interest.

However, as a general rule, if a process only occurs in vacuum via
quantum tunneling, then above some temperature it occurs much faster via
thermal activation.
(Chemistry and condensed matter physics are full of examples; annealing
of crystal defects, for instance.)  The same is true for baryon number
changing processes in the standard model, although the ``annealing
temperature'' is $O(100 {\rm GeV})$.  In fact, there is a phase
transition at $T_{\rm c} \sim 100$GeV in which the Higgs field loses its
condensate, and above this transition baryon number violation is
efficient \cite{KRS85,ArnoldMcLerran}.  It is quite possible that the
baryon number of the universe originated in a cosmological electroweak
phase transition, and in recent years this belief has driven the study
of the electroweak phase transition.

It takes two things for a cosmological electroweak phase transition to
generate the baryon asymmetry of the universe.  First, there needs to be
enough CP violation to generate at least the observed abundance of
baryons during the transition.  Second, there cannot be too
much ``annealing'' of the baryon number after the phase transition,
that is, baryon number violation must be inefficient enough after
the phase transition that a good fraction of the baryons survive to
the present day.  This is a condition on the phase transition's strength.

The minimal standard model fails both conditions badly
\cite{Gavela,KLRSresults}.  However, well motivated extensions, like the
MSSM with a light right scalar top, appear to be viable.  Recent studies of
baryon number production during the phase transition appear to show that
enough CP violation can hide in places with few low energy consequences 
to generate the observed abundance of baryon number, and maybe a little
more (for recent work see, for instance, \cite{Riotto,Rajagopal,CJK}).  
And the phase transition can be
stronger.  If the lightest scalar top is not very light, then
perturbation theory can reliably relate the phase transition in the MSSM
to the phase transition in the same effective theory used to study 
the standard model \cite{MSSMDR}, which has been well analyzed
numerically \cite{KLRSresults,KLRSSU2U1,Kripfganz}.
If the lightest scalar top is lighter still, the phase transition may
be stronger and more exotic \cite{Lainestop}.  This system can also be
studied by nonperturbative lattice techniques \cite{SU3SU2}.

At present, the weakest link in our knowledge of baryon erasure after
the phase transition is the relation between the strength of the phase
transition, now known nonperturbatively, and the efficiency of baryon
number violation after it is over, for which we have only a one loop
calculation \cite{ArnoldMcLerran,Carson,Baacke}.  We know
that the perturbation expansion at high temperature near the electroweak
phase transition cannot be viewed as an expansion in $\hbar$, but at
best as an expansion in a ratio of couplings.
We also know that the two loop corrections in the perturbative 
expansion for the strength of the phase transition aren't very small in
the ``interesting'' range of couplings where the baryon number violation
after the transition is close to the efficiency limit.  So it would be 
nice to actually know how good the one loop calculation is, or to
replace it with a fully nonperturbative investigation.

Very recently we
have proposed a nonperturbative method to determine the rate of baryon
number violation in the broken electroweak phase \cite{sphaleron1}.
This paper will fill in all the details left out in that paper.  Also,
the calculation there was incomplete; it did not include a measurement
of the ``dynamical prefactor,'' discussed below.  This paper will
complete this aspect of the calculation.  It will also 
discuss the importance to the broken phase sphaleron rate of
hard thermal loops, which can modify the dynamical prefactor.

For the impatient reader, we will present the basic ideas and the
results right now.  To find the sphaleron rate nonperturbatively, we
first define nonperturbatively a surface called the separatrix, 
which sits half way between distinct 
topological vacua.  Sphaleron transitions which permanently change
$N_{\rm CS}$ must cross this surface.  To find
the $N_{\rm CS}$ diffusion rate, we first compute the probability in the
canonical ensemble to be in a narrow band about the separatrix; then we 
compute the mean inverse time to cross the band.  The product is the
probability flux across the separatrix.  
Then we compute a ``dynamical prefactor,''
which tells what fraction of crossings lead to permanent resettling
about a different vacuum.  All three quantities can be computed
nonperturbatively on the lattice, using a combination of Monte-Carlo and
real time techniques.  Including strong hard thermal loop effects
modifies the dynamics in a way which lowers the dynamical prefactor, but
for realistic parameter values the effect is minor.
The $N_{\rm CS}$ diffusion constant is presented,
and compared to an analytic estimate based on the two loop effective
potential, in Table \ref{Table1}.

\begin{table}[t]
\centerline{\mbox{\begin{tabular}{|cc|c|c|c|}\hline
 & $(x \equiv \lambda / g^2)$ 
& $x = 0.047$ & $x = 0.039$ & $ x = 0.033 $ \\ \hline
&   $\phi(T_{\rm c}) / gT_{\rm c}$  &  1.360  
	& 1.568  &  1.789  \\ \cline{2-5}
``2 loop'' & $B \equiv g E_{\rm sph} / 4 \pi \phi$ & 1.643  &  1.633  
	&  1.626  \\ \cline{2-5}
perturbative& $E_{\rm sph} / T_{\rm c}$  &  28.08  &  32.20  
	&  36.55  \\ \cline{2-5}
 & $- \ln (\Gamma_d T_{\rm c}^{-4})$ &  22.27  &  25.39  
	&  28.82  \\ \cline{2-5}  
& $ - d \ln (\Gamma_d T^{-4}_{\rm c} ) / dy $  
	&  860  &  920  &  1000  \\ \hline
  &  $\phi(T_{\rm c}) /gT_{\rm c}$  &  $1.38 \pm 0.02$  
	&  $1.60 \pm 0.01$  & $1.82 \pm 0.03$   \\ \cline{2-5}
nonperturbative &  $- \ln ( \Gamma_d T_{\rm c}^{-4} )$ (excl. prefactor)  
	&  $24.7 \pm 0.4$  &  
	$28.3 \pm 0.4$  & $31.2 \pm 0.6$  \\ 
\cline{2-5}  &  $- \ln ( \Gamma_d T_{\rm c}^{-4} )$ (incl. prefactor)
	&  $25.9 \pm 0.5$  &  
	$29.5 \pm 0.5$  & $32.4 \pm 0.7$  \\  \hline
\end{tabular}}}
\caption{ \label{Table1}
Perturbation theory versus nonperturbative $\Gamma_d$.
Appearances of $T_{\rm c}^{-4}$ are really $(2.5 g^2 T_{\rm c})^{-4}$.
$x \equiv \lambda / g^2$ is the ratio of the Higgs self-coupling to the
gauge coupling, and $y \equiv m_H^2(T) / g^4 T^2$ is the dimensionless
Higgs mass squared.
The error bars for the nonperturbative $\phi_0$ are dominated by 
statistical errors
in the determined value of $T_{\rm c}$; errors in the nonperturbative
value of $\Gamma$ are statistical errors from the Monte-Carlo.  The last
row includes the nonunity dynamical prefactor in the rate and should be
taken as our most reliable estimate.}
\end{table}

The paper is structured as follows.  In Section \ref{ideasec}, we outline
the general idea of the calculation.  Section \ref{defNCSsec} defines
Chern-Simons number on the lattice, and the order parameter we will use,
which is very closely related.  It also discusses application of the
definition to the symmetric phase case.  Section \ref{Montesec} tells
how we go about things numerically.  Section \ref{Resultsec} presents
numerical results and compares them to a ``semi-two loop'' analytic
estimate, and to the erasure bound.  The last section concludes.
For readers who are allergic to details of numerical studies, we
recommend reading Section \ref{ideasec} carefully, and perhaps the first
subsection of Section \ref{Montesec}, and then skipping to
Section \ref{Resultsec}.

\section{Broken phase measurement: general idea}
\label{ideasec}

We want a technique for determining the $N_{\rm CS}$ diffusion constant
in the broken electroweak phase, where the rate is extremely small.  The
technique will be geared around the smallness of the rate and the fact
that the system in finite volume will spend almost all of its time in a
``neighborhood'' of a topological vacuum, in a sense to be made precise
below.  These assumptions can be checked {\it a posteriori}, and do not
constitute a real limit to the technique in the broken phase.  They will
fail in the symmetric phase or when the phase transition is very weak, 
but in that case we can apply real time techniques 
\cite{GrigRub,Ambjornetal,AmbKras,Moore1,Moore2,TangSmit,MooreTurok},
which can now produce quantitative results
\cite{slavepaper,AmbKras2,particles}.  (We should mention here that the
symmetric phase case is not completely settled; it has recently been
argued that there are logarithmic corrections to the parametric scaling
behavior \cite{Bodek_log}, which are however too small to be seen over
the noise and other systematics still present in \cite{particles}.)

\subsection{thermodynamic approximations}

Before we start to describe our approach to the calculation we will
specify the approximations to be made.

We treat the thermodynamics of
the standard model, or whatever extension is of interest, in the dimensional
reduction approximation \cite{oldDR,FKRS1,KLRS}, that is, as being well
approximated by a three dimensional, bosonic path integral with
parameters carefully matched to those of the full theory.  This is an
excellent approximation and we have no regrets in making it.  For a
study of corrections to this approximation in the present context,
see \cite{fermiondet}, which shows that the leading thermodynamic
effects not included in the dimensional reduction procedure have a
negligibly small effect on the sphaleron energy.

Conveniently, dimensional reduction is
equivalent to treating the theory's thermodynamics as equivalent to
those of the classical bosonic theory \cite{AmbKras}, with certain 
mass counterterms.  Similarly, we can treat the
theory's dynamics in a classical approximation.  This should be valid in
the infrared \cite{MooreTurok,Bodek_hbar}, with one serious
complication.  That is, the structure of divergent radiative corrections
to unequal time correlators is much more complicated than that for equal
time correlators.  For the equal time correlators, which are all that
matter to thermodynamics, the divergent radiative corrections are
mass squared corrections for the Higgs and $A_0$ fields, which can be
computed once and balanced by counterterms.  For unequal times the
linearly divergent radiative corrections, the hard thermal loops,
have a more complicated structure.  And they
can significantly change the dynamical behavior on time scales
longer than the inverse plasma frequency.  The modifications can be
important for the Chern-Simons number diffusion rate 
\cite{ArnoldYaffe,HuetSon,Son,Bodek_log}, as has recently been verified
numerically for the symmetric phase dynamics \cite{particles}. 
In the current context they will modify the ``dynamical prefactor'', but
they have little bearing on those parts of the calculation which are
thermodynamical.

We will also frequently make the approximation that the so called ``heavy''
degrees of freedom can be integrated out \cite{KLRS}, including both the
time component of the gauge fields and any
squarks present, so the theory reduces to an effective theory for the
minimal standard model \cite{MSSMDR}.  This theory is specified by two
parameters; $x \equiv \lambda / g^2$ 
the ratio of scalar to gauge self-couplings,
and $m_{\rm H}^2(T)$ the thermal Higgs mass squared Lagrangian
parameter.  $m_{\rm H}^2(T)$ is a monotone increasing function of $T$,
going from $O(g^2 T^2)$ at high temperatures to quite negative 
(symmetry breaking) in vacuum.  The phase transition occurs near 
where it is zero, at a critical temperature which in the context of 
dimensional reduction becomes a critical $m_{\rm H}^2(T)$.
At tree level, $\lambda /
g^2 = (m_{\rm H} / m_{\rm W})^2 / 8 $
but the radiative corrections are
important and the relation between $\lambda /g^2$ and the ratio of
physical zero temperature masses $m_{\rm H}/m_W$ is not
simple, especially in extensions to the standard model with new light
bosons.  Also note that, for instance, if the MSSM right scalar top is
too light, the reduction to an MSM like effective theory is not very
reliable; we should use an effective theory which contains the light
squark and the gluons.  Dropping heavy modes is not a necessary step for
using our technique, it is merely convenient to reduce the numerical
demands, which would be a few times larger if we include the $A_0$
fields and order 10 times larger if we include the squarks and QCD.  We 
will discuss how we think light squarks would change our results in
the conclusion.  We discuss the matter of integrating out the $A_0$
field in more detail in subsection \ref{algorithm_sec}.  
It is not always appropriate
to do so, and in particular we cannot when we are studying the influence
of hard thermal loops.

We do keep the $U(1)$ subgroup, which is often left out in electroweak
studies.  Its role in setting the sphaleron rate is probably almost
entirely due to its effect on the strength of the phase transition and
not a direct modification of the sphaleron, see \cite{Manton}, but the
numerical cost of including it is small enough that dropping it is
pointless.  We use $\tan^2 \Theta_{\rm W} = .32$, based on a 1 loop
match between vacuum $\overline{\rm MS}$ and 3-D thermal values using
results in \cite{KLRS}.

\subsection{the separatrix}

The idea of the separatrix between vacuum states is essential to our
technique.  Before introducing it, let us review what we expect the 
space of gauge-Higgs configurations to look like in the broken phase.
The space of three dimensional
gauge-Higgs configurations is periodic, with a discrete set of vacua.
To be more precise, we should consider the space of gauge-Higgs
configurations modulo (all) gauge transformations.  In this case all
vacua coincide\footnote{If the global topology of space is
multiply connected then Yang-Mills theory has
a connected manifold of inequivalent vacua corresponding to different
values for traces of certain noncontractible Wilson loops.  The toroidal
spaces we will consider are multiply connected; however, the (fundamental
representation) Higgs condensate lifts the degeneracy of the would be gauge 
vacua.  These complications will not be important for what we do.}, 
but the space is not simply
connected.  Since the index of the Dirac operator remembers when we go
around a noncontractible loop in configuration space, the relevant space
of physical configurations is the universal cover of the space of
configurations modulo (all) gauge transformations.  The universal cover 
has a discrete set of vacua
labeled by the index of the Dirac operator\footnote{The universal cover
is roughly the same as the space of configurations modulo small gauge
transformations.  But we prefer to think in terms of 
the universal cover of the space of configurations modulo all 
gauge transformations, because this is an
explicitly gauge invariant approach.}.  If the line connecting two vacua
in the universal cover projects to a winding 1 loop--or in more
conventional language, if two vacua differ by 1 in Chern-Simons
number--we will refer to them as neighboring vacua.

We expect that, in the
broken phase, almost all of the weight of the canonical ensemble lies in
states which are {\it in some sense} close to one of the vacua.  The
Hamiltonian evolution of a generic state in the ensemble will wander
around in the neighborhood of one vacuum for an exponentially long time
before it happens to make an excursion
far enough away that it crosses to being nearest another vacuum, which
it may find instead.  We expect that this is how $N_{\rm CS}$ diffusion
will occur.

\begin{figure}[t]
\centerline{\mbox{\psfig{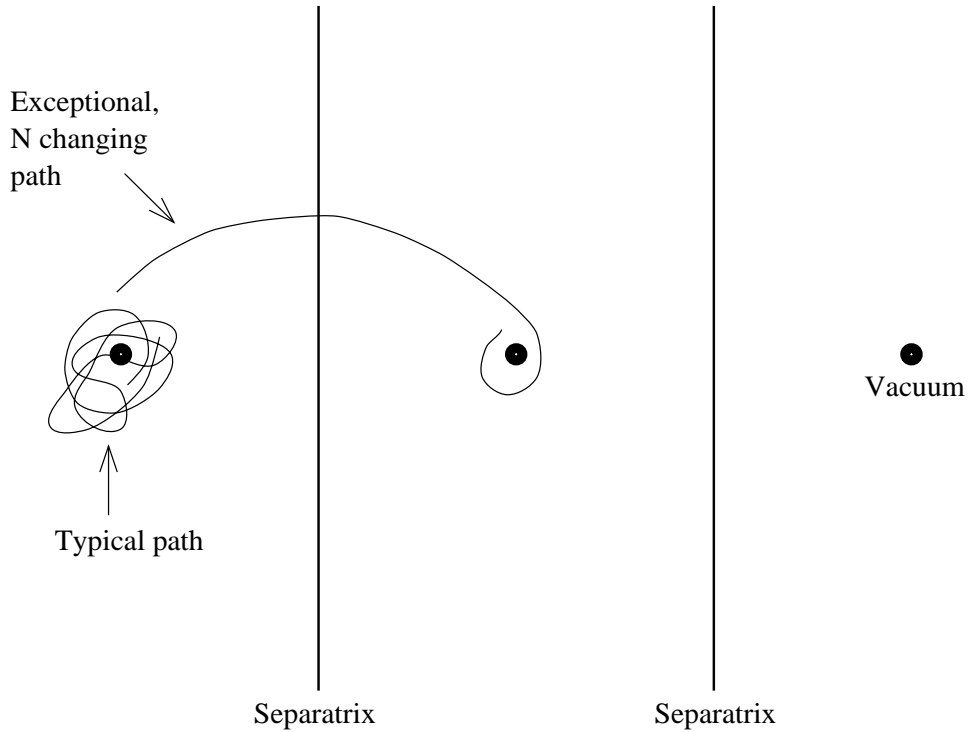}}}
\caption{\label{separatrix} Cartoon of periodic vacuum structure, 
separatrix, typical path which stays near a vacuum, and exceptional path
which crosses the separatrix and leads to permanent $N_{\rm CS}$ change}
\end{figure}

To determine the rate of $N_{\rm CS}$ diffusion, we draw a (codimension
1) surface separating one vacuum from its neighbor, halfway between the
minima in some sense, so that all of the well populated area near one
minimum falls on one side and all the well populated area near the other
minimum lies on the other side.  This surface is called the separatrix
between the vacua.  To cross from being near one vacuum to being near
another vacuum, a Hamiltonian trajectory must pass through the
separatrix dividing them.  We will assume that, 
after such a crossing, the trajectory
almost never promptly continues to and crosses the next separatrix, but
instead either settles around the new minimum for long enough that
ergodicity ``erases its memory,'' or turns around and returns to the
vacuum it started from.  Then the flux of probability of the thermal 
ensemble through the separatrix is an {\it upper bound} on the diffusion
rate for $N_{\rm CS}$.  It is an upper bound because of trajectories
which cross the separatrix, turn around, and return to the original
vacuum.  These lead to flux of probability through the separatrix, but
not to $N_{\rm CS}$ diffusion.
To get the true diffusion rate, we need to find not only
the flux through the separatrix, but the average number of
crossings of the separatrix per long term change from being near one
vacuum to being near another.  We will call the reciprocal of this, the
fraction of separatrix crossings which are associated with permanent
$N_{\rm CS}$ change, the ``dynamical prefactor''.
If we know both the flux of probability across separatrices, and the
dynamical prefactor, then we know the diffusion constant for $N_{\rm
CS}$; the diffusion constant is
\begin{equation}
\gamma_d \equiv \lim_{t \rightarrow \infty} 
	\frac{ \langle (N_{\rm CS}(t) - N_{\rm CS}(0))^2 \rangle}
	{t} = ({\rm prefactor}) \times ({\rm flux}) \, .
\end{equation}
What we really want is the diffusion constant per unit volume, 
$\Gamma_d \equiv \gamma_d / V$.  We should use a volume which
is large enough to prevent finite size systematics but small
enough that there are almost never two simultaneous
$N_{\rm CS}$ changing events in different places.

\begin{figure}[t]
\centerline{\mbox{\psfig{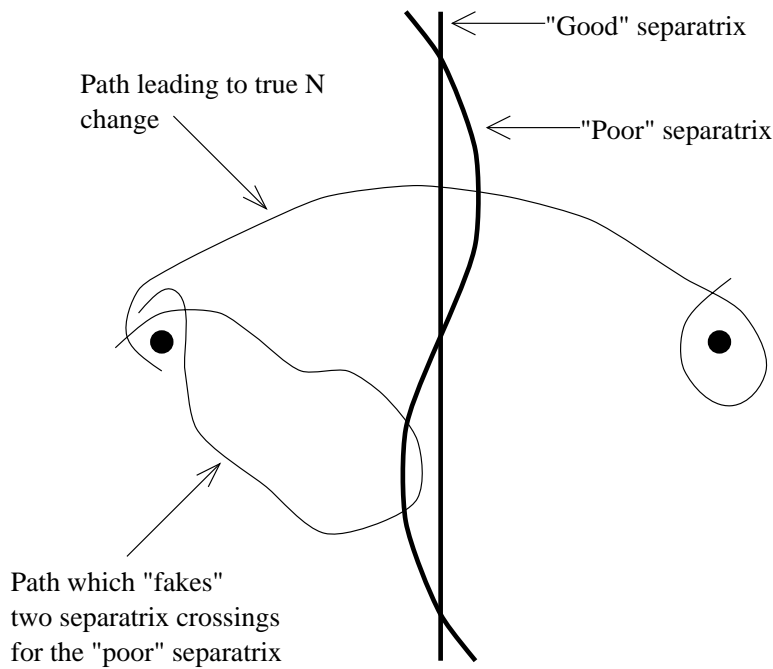}}}
\caption{\label{wrongseparatrix} Cartoon of how a poor choice of
separatrix can lead to overcounting the flux, and a small dynamical
prefactor.}
\end{figure}

A few points are in order here.  First, the assumption that there are no
prompt crossings of multiple separatrices is essential to the
calculation.  As we will see, it is also easy to test.  Second, if we
make a somewhat poor choice of the separatrix, so that there is some
place where it bends nearer to one vacuum than the other, then much of
the flux in that place will be of trajectories which double cross and
return to their starting vacuum.  The probability flux will be larger
than with a better definition of the separatrix.  However, the
dynamical prefactor will be smaller, since these extra crossings do not
lead to permanent $N_{\rm CS}$ change.  The rate we determine is
independent of exactly where we put the separatrix as long as the flux
across it is exponentially small, and as long as we make a complete
calculation, including the dynamical prefactor.  We illustrate this
point in Figure \ref{wrongseparatrix}.  In practice we should look for a
good choice of separatrix, since a poor choice of separatrix may
make it harder to get good statistics for $\Gamma$, as most
of the (numerical) effort will go towards studying trajectories which
double cross rather than ones which really change $N_{\rm CS}$
permanently.  

\subsection{calculation:  perturbation theory}

The existing perturbative calculations of the broken phase diffusion
constant for $N_{\rm CS}$ are along the lines of the
approach we just described.  To allow a perturbative calculation, they
make an additional assumption; that the separatrix is dominated by its
saddle point.  That is, they assume that gauge-Higgs
configurations on the separatrix look like perturbatively 
small excitations about a background field which is the lowest energy
point on the separatrix, which will be Klinkhamer
and Manton's sphaleron \cite{Manton}.  The choice of a definition of the
separatrix is then made perturbatively; a point is on the separatrix if
the excitation in the unstable direction of the sphaleron is zero.
The probability flux through the sphaleron can be computed in
perturbation theory by comparing the free energy of all excitations of
the sphaleron to the free energy of excitations about the naive vacuum,
with the frequency of the unstable mode serving to convert a probability
into a flux and the translational zero modes converting this into a flux
per unit volume.

The probability flux through the separatrix has been
computed in the above approximation 
at the one loop level \cite{Carson,Baacke}.  However, extending
the calculation beyond one loop raises severe technical problems.
The sphaleron is not a spatially homogeneous background field, so the
perturbative calculation must be done in real space with a numerically
determined spectrum of fluctuations.  The one loop calculation requires
finding this spectrum, but the two loop calculation involves overlap
integrals to compute the energy of their mutual interactions.
There are also conceptual problems, because one of
the fluctuation directions is unstable.  At one loop it is excised from
the sum over fluctuations; the other fluctuations set the probability to
be near the separatrix and it turns that probability into a flux through
the separatrix.  But it is not clear how to separate it from the other
modes at two loops.  These problems obstruct a systematic improvement of
the perturbative treatment.

There is also the problem of how to determine the dynamical prefactor
perturbatively.  Khlebnikov and Shaposhnikov argued that it should equal
1 \cite{KhlebShap}, but Arnold and McLerran made an estimate based on
Landau damping which suggests that it is quite a bit less than 1
\cite{ArnoldMcLerran}.  That argument has been more carefully developed
by Arnold, Son, and Yaffe, who claim that, so long as the Higgs
condensate gives a $W$ mass which is parametrically $m_W \sim g^2 T$, 
the prefactor should be
parametrically $O(\alpha_w)$ \cite{ArnoldYaffe}.  
Their argument has recently been tested in the symmetric phase
\cite{particles}.  No one has used their picture to get a quantitative
prediction of the dynamical prefactor within the context of
the perturbative calculation of the sphaleron rate, although this should
be possible in principle.

These limitations of the perturbative approach, together with the
generally spotty performance of perturbation theory for electroweak
phenomena at temperatures near the electroweak phase transition, 
motivate a fully nonperturbative attack on the problem.

\subsection{topology and the lattice}

The nonperturbative technique best suited to studying the sphaleron rate
is the lattice.  There is an apparent complication, though, which is
that topology cannot in general be well defined on the lattice; the
global structure of the space of three dimensional lattice gauge-Higgs 
configurations is different than in the
continuum, and in particular it is simply connected, so there are not
topologically distinct vacua.  However, if we make the lattice spacing
suitably small, then the thermal ensemble is completely dominated by
configurations in which {\it every} elementary plaquette is close to the
identity.  This means, roughly, that fields are perturbatively small at
the lattice scale. 
This subspace of the space of lattice configurations does
have the same topological structure as the continuum theory.  Hence, if
we excise a subspace of lattice configurations which carries an 
exponentially small weight, then we can talk about topology on the lattice.  

The physical meaning of this is as
follows.  The space of continuum configurations permits sphaleron-like
objects of arbitrarily small spatial extent.  They also exist on the
lattice, down to where their size is comparable to the lattice spacing;
but at this point it becomes unclear how to define a smoothly
interpolating continuum field, and the topological meaning is lost.
Such a configuration in the analogous 4 dimensional context is referred
to as an ``exceptional configuration.''
However, the energy of spatially small, sphaleron-like field configurations
rises linearly with inverse radius; so the Boltzmann suppression
of lattice scale sphalerons is enormous and they essentially never
occur.  As long as the ``genuine'' sphalerons we study are comfortably
larger than the lattice spacing, we have no problem.  And when the
genuine sphalerons we want to study are not comfortably larger than 
the lattice spacing, then obviously the lattice spacing is too coarse,
and we should use a finer lattice.

The situation here is much better than it is in the 4-D case considered
in QCD.  It is also true in 4 dimensions that if the gauge fields are
smooth enough, then topology is well defined
\cite{Luscher,Woit,Phillips}\footnote{in fact the 3-D case is a subset
of the 4-D one, since the topology we are talking about is the second
Chern class of a closed loop in 3-D configuration space, which is
equivalent to a periodic 4-D lattice configuration with the spacing in
the fourth dimension driven to zero.}.
However, in practice the fields may not generally be smooth enough.
Instantons just larger than the lattice spacing
typically do exist, because the instanton action is classically scale
independent, and so only varies logarithmically with instanton size.  
Making the lattice finer does not eliminate lattice spacing sized 
instantons (exceptional configurations) very quickly; in fact, because
their density goes as $\exp(-1/g^2)$ and $g^2$ varies logarithmically
with lattice spacing, their density declines as an algebraic power of
$a$.  In the 3-D context, though, the energy of a sphaleron of radius
$r$ goes as $1/(g^2 r)$ and the density of exceptional configurations 
varies with lattice spacing as $\exp(-({\rm coefficient})/(g^2 a T))$,
which falls off extremely rapidly as $a$ is made small.
This difference between the 3-D and 4-D cases
is because the 3-D theory is super-renormalizable; the coupling constant
is $g^2 T$, which is dimensionful.  Since $1/g^2$ appears in the
exponent for the rate of any nonperturbative phenomenon, and since it
must be accompanied by a $T$, on dimensional grounds a nonperturbative
phenomenon which occurs at the lattice spacing scale must proceed at a
rate which goes as 
$1/a$, leading to the exponentially fast rolloff in the density of
exceptional configurations as $a \rightarrow 0$.

\subsection{a ``nice'' nonperturbative choice for the separatrix}

Now that we have established that topological questions can have a
meaning on the lattice, we ask how to choose a separatrix which is
defined nonperturbatively and can be implemented on the lattice.

A definition we would prefer for the separatrix is the ``gradient flow''
definition, suggested in \cite{KhlebShap,Krishna,slave3}.  
It is defined in terms of gradient flow under the Hamiltonian.  The
Hamiltonian is a smooth function over the space of configurations, with
degenerate global minima at the vacua.  We believe that these are
the only local minima of the Hamiltonian in 3-D Yang-Mills Higgs theory,
although we do not know a proof.  Hence, following the direction of
steepest descent of the energy (gradient flow)
will lead, off a set of measure zero, to
a vacuum configuration.  A rigorous definition of ``the vacuum closest
to a configuration'' is the vacuum arrived at by such gradient flow,
which is also easily implemented on the lattice.  A very sensible
definition of the separatrix is then the boundary between the gradient
flow basins of attraction of two neighboring vacua (vacua with $N_{\rm
CS}$ differing by 1).  Equivalently,
it is the basin of attraction of the saddle point which sits between the
two vacua, i.e. the sphaleron.  If we mod out by all gauge transformations
the two vacua are equivalent, but the separatrix can still be defined as
the
surface where two infinitesimally separated configurations on opposite
sides will have macroscopically different gradient flow paths which,
when spliced together, form a noncontractible loop.

Alternatively we could define the separatrix just in
terms of the Yang-Mills field (connection) and the Yang-Mills term in
the Hamiltonian.  This choice has the added benefit that, as the
configuration gradient flows under the Yang-Mills Hamiltonian in 3
dimensions, it becomes exceedingly smooth (meaning that all gauge
invariant local measurables are slowly varying and the energy density
is very small).  Also, leaving out the Higgs fields evades the
complication that the Higgs mass squared is renormalized by ultraviolet
thermal excitations, which change during the gradient flow.  
The Yang-Mills gradient flow separatrix may not coincide with the
Yang-Mills Higgs separatrix (which is not defined until we decide how
to deal with the renormalization of the Higgs mass term in the
Lagrangian).  But if we perform a complete calculation, including the
dynamical prefactor, then the exact choice of separatrix should not
matter, as long as crossings are exponentially rare.  Of course, a poor
choice will make the calculation inefficient, since most crossings will
not be associated with topology change.  But we do not expect the
Yang-Mills gradient flow separatrix to be a poor choice, and we will be
able to check this belief when we study the dynamical prefactor.

\subsection{approach to computation}

Now, we will outline how to compute the flux and the dynamical
prefactor.  To do so, we need more than just a definition of a
separatrix.  We need an order parameter $N$ 
which takes a special value, say
$N=1/2$, on the separatrix, and is smaller on one side and larger on the
other (say, going from $N=0$ at one vacuum to $N=1$ at the other).
Assume that we have such an $N$.

First, we should 
measure the probability over the canonical ensemble that $N$ is 
within some small tolerance $\epsilon/2$ of $(\rm integer +) 1/2$.  
This gives the
probability to be very near a separatrix.  Since the probability to be
close to the separatrix is (expected to be) exponentially small, we will
need to use a multicanonical reweighting technique to sample here
accurately.  The basic idea is presented early in Section
\ref{Montesec}.

Then we need to know the mean inverse time for crossing 
this narrow region, to turn the probability into a flux.
This is $1/\epsilon$ times $\langle |dN/dt| \rangle$, 
the mean of the absolute value
of the time derivative of $N$, where the averaging is over the ensemble
restricted to the narrow band about the separatrix.  
We determine this by taking a
canonical sample of states with $| N - 1/2 | < \epsilon/2$, drawing
momenta for each from the thermal ensemble\footnote{It is essential here
that the phase space is the tangent bundle of the configuration space,
and that the thermal probability distribution is a product of a
configuration space function and a (Gaussian) function over momentum
space.}, and performing a short segment of Hamiltonian evolution,
comparing $N$ with its value a short Hamiltonian 
time later.  Thus, we can determine the flux.

To determine the dynamical prefactor, we take a canonically weighted 
sample of
configurations at the separatrix and choose momenta for each, just as we
do to find $\langle |dN/dt| \rangle$.  Then we
perform the Hamiltonian evolution both forward and backward in time, 
until the forward and backward histories both settle into the
neighborhood of a vacuum.  We count
how many times the Hamiltonian trajectory crosses the
separatrix before it settles semipermanently about a minimum.  The
prefactor is
\begin{equation}
{\rm Prefactor} = \sum_{\rm sample} \frac{1}{\rm number \; of \;
	crossings} \times \left\{ \begin{array}{cl} 
	1\;\;\; & {\rm final \; vacuum} 
	\neq {\rm initial \; vacuum} \\ 0 \;\;\; & 
	{\rm  final \; vacuum} = {\rm initial \; vacuum} \; . \\
	\end{array} \right.
\end{equation}
This is not the same as adding up the number of times the final vacuum
differed from the initial one and dividing by the number of crossings of
the separatrix.  The reason is that the latter overcounts trajectories
with many crossings, since the ensemble samples them more often than
trajectories with fewer crossings.

This is our recipe; given an order parameter $N$, we can determine both the
flux of probability through the separatrix, and the dynamical prefactor,
and hence $\Gamma_d$.

It remains to choose an order parameter.  We want one such that the
$N=1/2$ separatrix will be either the same as or quite close to the
``good'' choice of the gradient flow separatrix.
Naively, we expect a good $N$
to be the Chern-Simons number, $N_{\rm CS}$.  Actually, this is not
quite right, as we discuss at some length in the next section.

\section{Defining lattice $N_{\rm CS}$}
\label{defNCSsec}

We now have an idea for a nonperturbative definition of the separatrix,
but we need an order parameter which varies from 0 to 1 as it ranges
between vacua and equals 1/2 on a surface close to the Yang-Mills
gradient separatrix.  The literature generally considers the
Chern-Simons number $N_{\rm CS}$ to be the best choice for this
\cite{KhlebShap}, so we will discuss how to define $N_{\rm
CS}$ on the lattice; then we will define a different but closely related
order parameter $N$, which gives a separatrix much closer to the
gradient flow separatrix than $N_{\rm CS}$ does, and which we will 
actually use in the calculation of the broken phase diffusion rate.

\subsection{the definition}

Chern-Simons number should be defined as some real valued
function over the space of three dimensional gauge connections.
Let us review some properties which $N_{\rm CS}$ would have in the
continuum, and see how closely we can preserve them on the lattice.

In the continuum, $N_{\rm CS}$, defined on the space of gauge field
configurations modulo {\em all} gauge transformations, has 
the following properties:
\begin{enumerate}
\item   $N_{\rm CS}$ should be a continuous, multiply valued function, 
	with the values separated by 1.  
\item   The multiple values correspond to the projections of different 
	points on the universal cover, where $N_{\rm CS}$ 
	is single valued.  A noncontractible loop in the configuration 
	space lifts to a line segment in the cover space, 
	and the difference in $N_{\rm CS}$ between
	the endpoints is the winding number of the loop.  (A more 
	conventional way of saying this is that $N_{\rm CS}$ differs 
	between two gauge copies by the winding number of the gauge
	transformation between them.  But we prefer the above, gauge
	invariant statement.)
\item	A vacuum configuration has $N_{\rm CS}$ modulo 1 
	equal zero.
\item   Consider a path in configuration space parameterized by $\tau$.
	The difference in $N_{\rm CS}$ between the beginning and end
	of the path should be
\begin{equation}
\Delta N_{\rm CS} = \frac{g^2}{16 \pi^2} \int d\tau \int d^3 x
	\epsilon_{ijk} F^a_{ij} (D_{\tau} A_k)^a \, .
\label{intalongpath}
\end{equation}
	This is the same as saying that 
	$N_{\rm CS}$ is the indefinite integral of 
\begin{equation} 
\frac{g^2}{32 \pi^2} \int d^3 x F^a_{\mu \nu} \tilde{F}^a_{\mu \nu} 
	= \frac{g^2}{8 \pi^2} \int d^3 x E^a_i B^a_i \, .
\end{equation}
\end{enumerate}

How many of these properties can we preserve on the lattice?  Not all of
them; as we pointed out in \cite{Moore1}, there is no local 
operator definition of $E^a_i B^a_i$ which is a total derivative.  We
also argue there that there are severe difficulties satisfying 2.;
but this is because we demanded a singly valued, non gauge 
invariant definition of $N_{\rm CS}$.  
In fact we can present a lattice definition
of $N_{\rm CS}$ which satisfies everything but 4. and the continuity
requirement, provided we restrict to gauge fields which are smooth enough, 
in the sense of the last section, so that 2. makes sense.

\begin{figure}[t]
\centerline{\mbox{\psfig{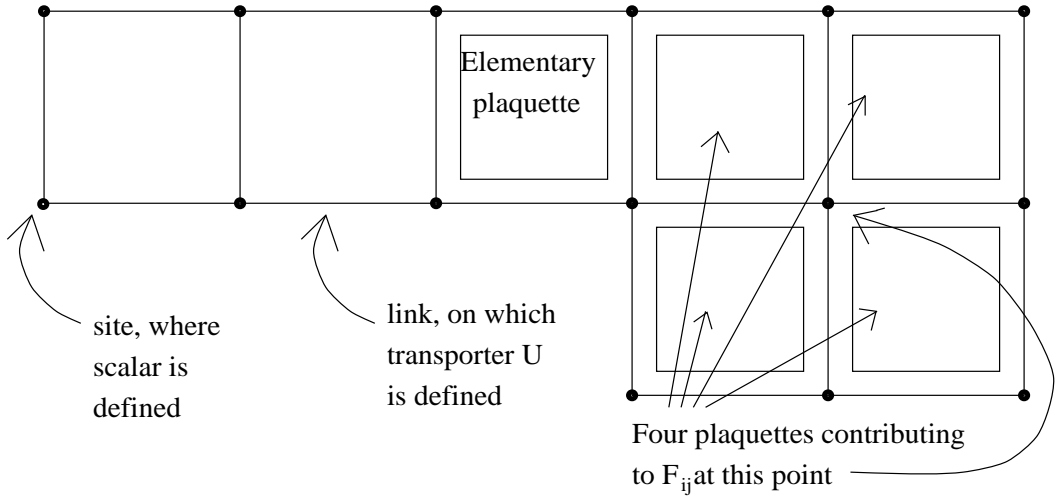}}}
\caption{\label{fig1b} Illustration of the pieces which make up a
lattice gauge theory.}
\end{figure}

Before constructing a lattice definition of $N_{\rm CS}$, we
remind the reader how the lattice fields are defined.  On a
lattice, scalar fields are only defined at a discrete set of points, the
lattice sites.
The gauge field should be a connection, that is, it should be a rule
which tells how to parallel transport fields along paths.  We allow a
path on the lattice to consist of a series of straight lines between
nearest neighbor lattice sites.  The connection is then defined by
associating a group element $U \in$SU(2) with 
each of these elementary straight lines
(referred to as the links of the lattice).  A small closed path, or the
product of the $U$ around the path, is called a plaquette;
the $1 \times 1$ square is the elementary plaquette.  When we do not
specify the shape of a plaquette we mean an elementary plaquette.  The
product of the $U$ around a plaquette 
(written $U_{\Box}$, often just referred to as
``a plaquette'') will not in general be the identity; its failure, a
curvature in the connection, is a field strength.

It is important that the field strength is not associated with a site of
the lattice, but with a plaquette, which sits in between sites.  Then,
to make a lattice implementation of $\int \epsilon^{\mu \nu \alpha \beta}
F_{\mu \nu} F_{\alpha \beta}$, we will have to do some averaging.  The
argument of the integral is a pseudoscalar and should perhaps be defined
at the lattice sites; but the field strength is associated with a
plaquette.  To preserve cubic symmetry, $F_{\mu \nu}$ at a site will
have to be the average over the four plaquettes in the $\mu,\nu$ plane
which touch the site.\footnote{We are thinking of a path through
configuration space as a 4 dimensional lattice, where the fourth
dimension is the path parameter, taken as a discrete rather than a
continuous variable.  Successive 3 dimensional slices are configurations
along the path.  In a numerical setting, a path through the 
configuration space will always look like this, though making the
spacing in the 4'th direction arbitrarily small restores the continuity
of the path.}  Because of this averaging process over things
which do not live quite at a lattice site, the resulting lattice
definition of $\epsilon^{\mu \nu \alpha \beta}
F_{\mu \nu} F_{\alpha \beta}$ will not be a total derivative; and we
cannot fix this problem by going to fancier definitions involving
weighted averages of plaquettes of various shapes \cite{Moore1}.  
The problem is that the continuum proof that $F \tilde{F}$ 
is a total derivative relied on
continuity of the fields, and this continuity is lost on the lattice.
This makes it impossible to satisfy 4.  Note, however, that the lattice
definition of $F_{\mu \nu} \tilde{F}_{\mu \nu}$ is gauge invariant.

However, when the gauge fields are weak at the lattice scale
(meaning that the elementary plaquettes are close to 
the identity) and slowly varying (meaning that the departure 
from the identity is nearly the
same between a plaquette and the parallel transport of a nearby
plaquette in the same plane), then the lattice definition of $E^a_i
B^a_i$ is approximately a total derivative.  Further, we can pick a
definition of $N_{\rm CS}$ so that 4. is satisfied 
at least in one particular special
direction.  We choose it to be true in the cooling direction, that is,
the direction of energy gradient flow.  Here the 
energy of the lattice field is given by the standard Kogut-Susskind
Hamiltonian \cite{Kogut}, which is defined up to a multiplicative 
factor as 
\begin{equation}
\label{KogutSusskind}
H_{\rm KS} ( U ) \propto \sum_{\rm plaquettes \; U_{\Box}} \left( 
	1 - \frac{1}{2} {\rm Tr} U_{\Box} \right) \, .
\end{equation}
The gradient of $H_{\rm KS}$ is to be understood in terms of the metric
of the configuration space, which is the product of the Haar measure
over each link matrix $U$.

We will call a path which follows the steepest descent of
$H_{\rm KS}$ a cooling path, and we parameterize it with a
cooling time $\tau$, defined as $d\tau = d{\rm (path \; length)} / 
[dH_{\rm KS}/d{\rm (path \; length)}]$.  The gauge fields evolve along
the path according to \cite{AmbKras2}
\begin{equation}
\frac{\partial U}{\partial \tau} = - D^\alpha U D^\alpha H_{\rm KS} \, ,
\end{equation}
where $D^\alpha$ is the left acting covariant derivative, 
$D^\alpha U = i \tau^\alpha U$, and $D^\alpha H_{\rm KS}$ is $H_{\rm
KS}$ with $U$ replaced with $D^\alpha U$.  This is the gauge invariant
lattice implementation of the continuum evolution
\begin{equation}
\frac{ \partial A_i^a(x)}{\partial \tau} = - \frac{\partial H}
	{\partial A_i^a(x)} \, .
\end{equation}
$\tau$ has dimensions of length squared.

\begin{figure}[t]
\centerline{\mbox{\psfig{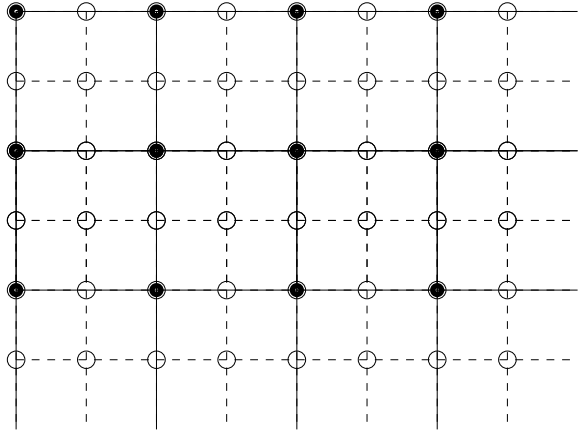}}}
\caption{ \label{Fig2}
	Original lattice and the coarsened version, which is just the
solid lines and filled vertices.  The link matrix between two vertices
of the coarsened lattice is the product of the matrices on the two
original links which make it up.}
\end{figure}

We are defining $N_{\rm CS}$ so that condition 4. is true along the
cooling path.  Since cooling eventually leads to the vacuum off a set of
measure zero, and since $N_{\rm CS}$ of the vacuum is by definition an
integer, we get
\begin{equation}
\label{Def_of_NCS}
N_{\rm CS} \equiv \, {\rm integer} \, - \int_0^{\infty} d\tau \frac{g^2}{8
	\pi^2} \int d^3 x E_i^a B_i^a \, ,
\end{equation}
where we mean the lattice definition of $E_i^a B_i^a$ which we have
described, and which is written down in \cite{AmbKras}.
This definition of $N_{\rm CS}$
is nice because the cooling path leads most quickly
towards configurations with weak, slowly varying gauge fields, so the
definition is minimally contaminated by the problems with the lattice
definition of $E_i^a B_i^a$ which we have discussed.

Performing the integral in
Eq. (\ref{Def_of_NCS}) takes an enormous numerical effort.
Cooling the fields (following the cooling path) stably requires using a
$\tau$ step size of less than $a^2/6$ ($a$ the lattice spacing), 
or else the most ultraviolet
excitations become unstable.\footnote{We improve
on this marginally by using alternately larger and smaller step sizes,
analogous to what we did in \protect{\cite{Moore1}} when quenching the
$E$ fields to enforce Gauss' Law.}  But sufficient cooling may demand
going to $\tau$ of order $1000 a^2$.  However, the cooling is by far the
most efficient at removing ultraviolet excitations, and 
already by $\tau > a^2$ the fields are slowly varying.  After this much
cooling we 
lose almost no information if we drop some UV degrees of freedom by
setting up a coarsened lattice.  Define an even site as a site where all
three coordinates are even numbers.  In a scalar field
theory, we would coarsen by dropping out
all the lattice sites which are not even, leaving a lattice half as many
points across in each direction.  In a gauge theory, we
also need to define the connections between the sites of the coarsened
lattice; we define the connection between two neighboring 
even sites as the product
of the two connections along the straight line between them.  We
illustrate the idea in Figure \ref{Fig2}.  The remainder of the
cooling then proceeds $2^5$ times faster, $2^3$ because the lattice
is smaller and $2^2$ because we can use a $\tau$ step size which is
larger in physical units.  

Of course we must check
that this procedure produces the same answer as we get by not
coarsening.  But if we use an $O(a^2)$ improved definition of $E \cdot
B$, for instance the one we developed in \cite{Moore2}, then this is not
a problem at all.  The value of $N_{\rm CS}$ gets rescaled by an amount
typically less than $1 \%$ and has an amount of noise added to it which
is typically even smaller.  For lattices more than about $28$ sites
across, we can even safely perform a second stage of coarsening after
performing several coolings on the once coarsened configuration.

\begin{figure}[t]
\centerline{\mbox{\psfig{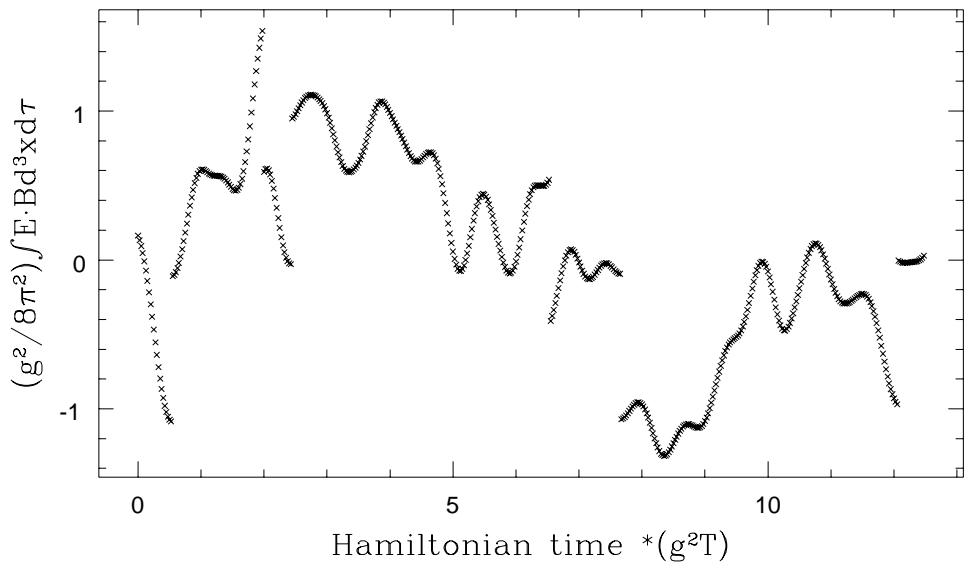}}}
\caption{ $(g^2/8\pi^2)\int E_i^a B_i^a d^3 x d \tau$ 
for a series of points on a Hamiltonian trajectory.  It is clear where
to adjust the integer part of Eq. (\protect{\ref{Def_of_NCS}}) to keep
$N_{\rm CS}$ (approximately) continuous. \label{jump_fig}}
\end{figure}

\subsection{application to the symmetric phase}

As an application we discuss how to use this definition of $N_{\rm CS}$
to study $N_{\rm CS}$ diffusion in the symmetric phase, or pure
Yang-Mills theory.  In this case we want to track $N_{\rm CS}$ along the
projection into configuration space of a Hamiltonian trajectory in phase
space.  That is, generating a Hamiltonian trajectory will give us a
closely spaced series of points in configuration space, and we must
associate a single valued function $N_{\rm CS}$ with this series.  Our
definition says that we should measure $N_{\rm CS}$ at each point by
performing the integral in Eq. (\ref{Def_of_NCS}), that is we must cool
the configuration at each point along the Hamiltonian trajectory to the
vacuum.  We also have to choose the integer part of
Eq. (\ref{Def_of_NCS}) somehow.  At certain points, the value of the
integral will abruptly jump by almost an integer.  This happens whenever
the Hamiltonian trajectory crosses a gradient flow separatrix.
We should choose the integer part of $N_{\rm CS}$ to minimize
the magnitude of the jump in $N_{\rm CS}$
at the separatrix.  We illustrate with an
example of real data from a simulation of Yang-Mills theory in
Fig. \ref{jump_fig}.  Note that the discontinuities in $N_{\rm CS}$ do
not necessarily occur when $N_{\rm CS}$ is near $\pm 1/2$; we will discuss
this more in the next subsection.

\begin{figure}[t]
\centerline{\mbox{\psfig{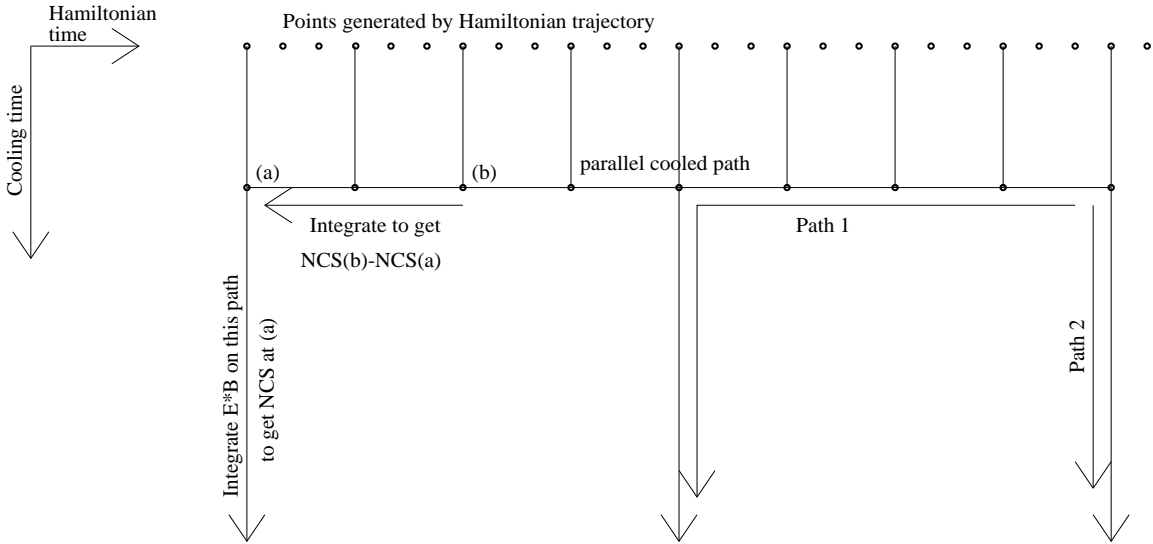}}}
\caption{(Time,cooling time) plane, with curves used to track $N_{\rm CS}$
for a Hamiltonian trajectory.  Every few Hamiltonian updates, we
construct a cooled copy of the configuration, giving a parallel cooled
path.  Every few points on that path, we measure $N_{\rm CS}$ directly,
and we fill in between by integrating $E\cdot B$ along the parallel
cooled path.  As long as the integrals along paths 1 and 2
always agree modulo an integer, within a small tolerance, the technique
is topological. \label{calibrate}}
\end{figure}

In practice, even with lattice coarsening, the numerical costs of
cooling every configuration are unbearable, but we can do better.  We
show how in Fig. \ref{calibrate}.  Every few steps, we cool a little, to
a depth of $\tau \sim a^2$, and we thereby construct a cooled image of
the Hamiltonian path, a technique explored by Ambj{\o}rn and Krasnitz 
\cite{AmbKras2}.  We measure $N_{\rm CS}$
using Eq. (\ref{Def_of_NCS}) at occasional points along this path,
interpolating $N_{\rm CS}$ in between by integrating $E_i^a B_i^a$ along
the cooled image of the Hamiltonian path.  The cooling has eliminated
most of the UV excitations, so $E_i^a B_i^a$ along the cooled path 
is close to behaving as a total derivative, especially if we use an 
$O(a^2)$ improved definition of $E^a_i B^a_i$.  The interpolated value 
of $N_{\rm CS}$ is therefore almost what we would get by using the 
direct definition at each time.  We only need it to
be close enough to determine the integer part of Eq. (\ref{Def_of_NCS})
unambiguously, which we can if the value we get by integrating on 
path 1 in Fig. \ref{calibrate} and the value we get by integrating on 
path 2 in that figure differ by an integer plus a small remainder.  We
can think of this remainder as a calibration of the integration along
the cooled path, so the approach is a ``calibrated cooling method'', 
with the occasional cooling paths to the vacuum recalibrating the method 
of Ambj{\o}rn and Krasnitz \cite{AmbKras2} to make it topological.

\begin{figure}[t]
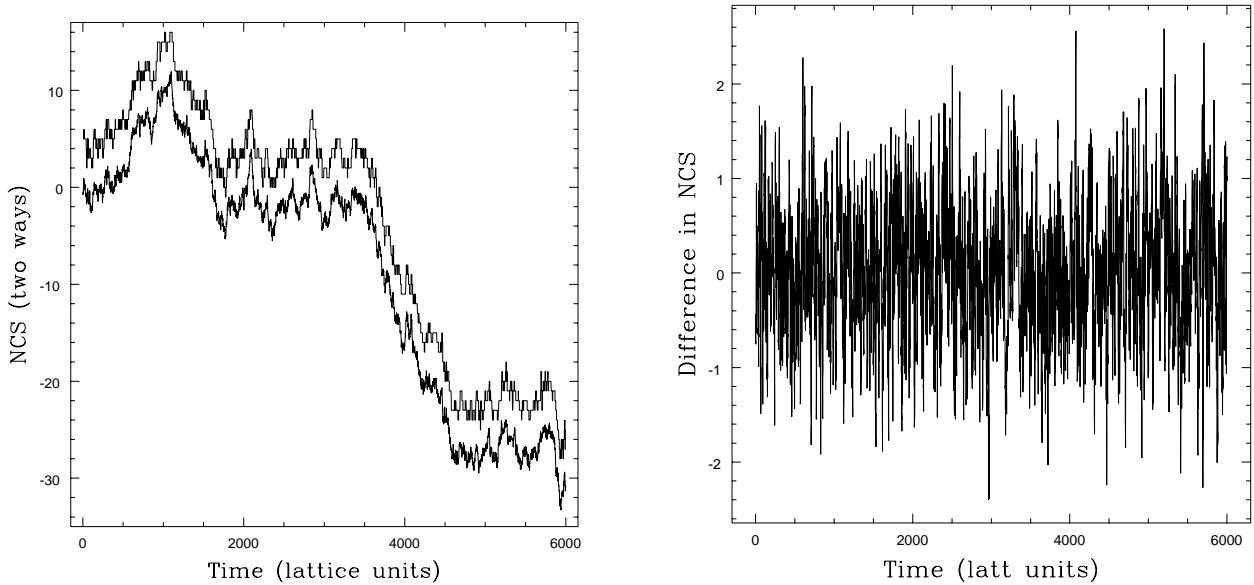

\centerline{\mbox{\psfig{file=wind.epsi,width=3in}} \hspace{0.4in}
\mbox{\psfig{file=diff.epsi,width=3in}}}
\caption{Left:  $N_{\rm CS}$ measured by the slave field (upper curve) and
``calibrated cooling method'' (lower curve).  Right:  the difference
between the curves (note scale).  
The difference is small and spectrally white,
so the methods are in good agreement. \label{compareNCS}} 
\end{figure}

We compared this approach of measuring $N_{\rm CS}$ to the ``slave
field'' topological method \cite{slavepaper}, by evolving Yang-Mills theory
on a $24^3$ grid at $a = 1/(2g^2T)$ ($\beta_L = 8$) for a total time of 
$6000 a$, tracking $N_{\rm CS}$ by each technique.  For the technique we
just described, we constructed a cooled 
image path with one point every $a/5$ time.  The
cooling depth to this path was $5 a^2/8$, and we calibrated by cooling to
the vacuum every $2a$ time.  With this lattice spacing and this
frequency of calibrating, the largest remainder we observed was $0.2$ and
the typical absolute value was less than $0.05$.  We present the
results for $N_{\rm CS}$ in 
Fig. \ref{compareNCS}.  We have offset $N_{\rm CS}$ measured
by the slave field method by 5 to keep the curves from lying on
top of each other.  The agreement is outstanding, and the difference in
the determined values of $N_{\rm CS}$ is white on long time scales.

To explain why the two methods have a white noise difference, we review
briefly how the slave field method works.  It tries to keep track of the
``integer'' part of Eq. (\ref{Def_of_NCS}) by assuming that the cooling
path will end in a vacuum which has winding number zero in Coulomb
gauge, and then adding up the number of large gauge transformations
required to keep the system in Coulomb gauge during the Hamiltonian
trajectory.  However, this ignores the contribution to
Eq. (\ref{Def_of_NCS}) from the integral, that is, the difference
between $N_{\rm CS}$ of the configuration and of the vacuum it cools to.
Also, the algorithm used to find Coulomb gauge sometimes gets trapped
temporarily in a Gribov copy with a different winding number.  But
neither difference between the methods will grow without limit in time,
so the difference between the two measurement methods for $N_{\rm CS}$
is white on long time scales.  Hence, the derived diffusion constant
will be the same within errors (caused by the white noise difference).
Indeed, when we used the technique of
\cite{slavepaper} to extract $\Gamma_d$ from each trajectory, the two
methods of tracking $N_{\rm CS}$ gave the same answer within error
($\gamma_d = .0515 \pm .0077 a^{-1}$ for the new method, 
versus $\gamma_d = .0516 \pm .0082 a^{-1}$ for the slave method).

As an aside, we mention that Hetrick and de Forcrand have used the method
of cooling to resolve the Gribov problem, by defining Coulomb gauge (or,
in their 4 dimensional context, Landau gauge) as the gauge in which the
vacuum configuration arrived at via cooling has the minimum value of
$\int A^2$ \cite{Hetrick}.

\subsection{modifications for measuring $\Gamma_d$ in the broken phase}

The definition of $N_{\rm CS}$ which we just presented more accurately
reproduces the continuum meaning of $N_{\rm CS}$ than any other we know.
But we don't actually want all of the attributes of the continuum
meaning of $N_{\rm CS}$ if we want to measure $\Gamma_d$ in the broken
phase by the separatrix method.
The reason is that $N_{\rm CS}$ is only directly a measure of topology
for vacuum configurations.  There are contributions to $N_{\rm CS}$ in
excited states which are uncorrelated to topology; for instance, $N_{\rm
CS}$ does not vanish in abelian gauge theory, even though that theory
has no $\pi_3$ topological sectors.  In the continuum abelian theory,
$N_{\rm CS}$ is given by
\begin{equation}
N_{\rm CS} = \frac{ g^2 }{32 \pi^2} \int d^3 x \epsilon_{ijk} 
	 F_{ij} A_k \, ,
\label{naiveNCS}
\end{equation}
and the mean square value of $N_{\rm CS}$ is
\begin{equation}
\langle N^2_{\rm CS} \rangle = \frac{g^4}{1024 \pi^4} \int d^3 x d^3 y
	\epsilon_{ijk} \epsilon_{lmn} \langle F_{ij}(x) A_k(x) F_{lm}(y)
	A_n ( y ) \rangle \, .
\end{equation}
Using Wick's theorem and the momentum representation of the propagator,
in a general covariant gauge, this becomes
\begin{eqnarray}
\langle N^2_{\rm CS} \rangle & = & \frac{g^4}{1024 \pi^4} \int d^3 x d^3 y
	\int \frac{d^3 p d^3 q}{(2 \pi)^6} e^{i (p+q) \cdot (x-y)}
	\epsilon_{ijk} \epsilon_{lmn} T^2 \times \nonumber \\ &&
	\quad \left[
	\frac{4 p_i p_l}{p^2 q^2} \left( \delta_{jm} + (\alpha - 1)
	\frac{ p_j p_m}{p^2} \right) \left( \delta_{kn} + ( \alpha - 1)
	\frac{q_k q_n}{q^2} \right)
	+ ( m \leftrightarrow n) \right] 
	\nonumber \\
& = & \frac{g^4 T^2 V}{64 \pi^4} \int \frac{d^3 p}
	{ (2 \pi)^3 } \frac{p^2}{(p^2)^2} \, ,
\label{UVdiv}
\end{eqnarray}
so $N_{\rm CS}$ will be Gaussian distributed with a linearly divergent 
variance.  On the
lattice, the UV divergence will be cut off by the lattice scale; the
coefficient was found by Amjorn and Krasnitz \cite{AmbKras} and is
\begin{equation}
\langle N^2_{\rm CS} \rangle = (1.44 \times 10^{-5}) g^4 V T^2 / a \, .
\label{UVdiv_num}
\end{equation}

The divergence occurs because, while the energy cost of storing $N_{\rm
CS}$ in a UV mode grows linearly with $p$, the number of available
states in which to store $N_{\rm CS}$ grows faster; entropy wins over
energy.  The same thing happens in SU(2) theory, because $N_{\rm CS}$
also contains the $\epsilon_{ijk} F_{ij} A_k$ term.  The coefficient of
the divergence in SU(2) is larger by 3, the dimension of the group.  Since
Yang-Mills theory in 3-D is super-renormalizable, the UV decouples from
the IR, where the genuine topology changing physics occurs, so this UV
divergent, Gaussian contribution will appear as an additive correction
to the IR contribution to $N_{\rm CS}$.  That is, to reasonable accuracy
we can think of $N_{\rm CS}$, defined in Eq. (\ref{Def_of_NCS}), as
$N_{\rm CS} = N_{\rm CS}^{\rm IR} + N_{\rm CS}^{\rm UV}$, 
where topological information is in $N_{\rm CS}^{\rm IR}$, and $N_{\rm
CS}^{\rm UV}$ is independent of $N_{\rm CS}^{\rm IR}$ and Gaussian
distributed.  

Given a single IR field configuration with some particular value of
$N_{\rm CS}^{\rm IR}$, different realizations of the UV excitations on
top of the IR fields will then give a distribution of values of $N_{\rm
CS}$; so the probability that $N_{\rm CS}$ will have a particular value
$x$, $P_{\rm NCS}(x)$, will be
\begin{equation}
P_{\rm NCS}(x) = \int_{\rm IR \; configs} \int_{\rm UV \; excit.}
	\delta(x - N_{\rm CS}^{\rm IR} - N_{\rm CS}^{\rm UV} ) \, ,
\end{equation}
or, defining $P_{\rm NCS}^{\rm IR}(x)$ and $P_{\rm NCS}^{\rm UV}(x)$ to
be the probability distributions for the IR and UV components of $N_{\rm
CS}$, 
\begin{equation}
P_{\rm NCS}(x) = \int dy \int dz P_{\rm NCS}^{\rm IR}(y) 
	 P_{\rm NCS}^{\rm UV}(z) \delta(x-y-z) = 
	\int dy  P_{\rm NCS}^{\rm IR}(y)  P_{\rm NCS}^{\rm UV}(x-y) \, .
\end{equation}
The probability distribution for $N_{\rm CS}$ will be a convolution of
the interesting IR distribution and a Gaussian noise distribution.
Convolving any periodic function with a Gaussian always degrades
contrasts in that periodic function, enhancing
the probability to be near the $N_{\rm CS}=1/2$ separatrix relative to
the case of no UV noise.  The distortion will be small only if
\begin{equation}
\langle (N_{\rm CS}^{\rm UV})^2 \rangle \ll \left. \frac{d^2 \ln 
	P_{\rm NCS}^{\rm IR}(x)}{dx^2} \right|_{x=1/2} \, .
\end{equation}

Later, we will use a $40^3$ lattice with $a = 2/(5 g^2 T)$.  Multiplying
Eq. (\ref{UVdiv_num}) by the group factor of 3 and plugging in numbers,
$\langle (N_{\rm CS}^{\rm UV})^2 \rangle = 0.44$ for such a lattice,
which is too big.  So defining $N_{\rm CS}$ by Eq. (\ref{Def_of_NCS})
will not do.

Another way to state the above is that the separatrix one gets from the
condition $N_{\rm CS} = 1/2$ is sensitive to ultraviolet excitations,
which makes it ``all wiggly''; it will have lots of ``fingers'' which
stick out towards one or the other topological vacuum, and the problems
we discussed in Section \ref{ideasec}, of there being many crossings
of the separatrix which don't have to do with permanent $N_{\rm CS}$
change, will be severe.

The problem is that $N_{\rm CS} = 1/2$ is not particularly similar to
the ``good'' gradient flow definition of the separatrix.  We want
an order parameter which is close to $N=1/2$ on the gradient flow 
separatrix.  The $N=1/2$ separatrix does not need to correspond exactly
with the gradient flow definition; is sufficient if the distribution 
of values of $N$ on the gradient flow separatrix is narrow, preferably 
narrower
than $d^2 P_{\rm N}(x)/dx^2$.  A slight change to Eq. (\ref{Def_of_NCS})
will do the trick; define
\begin{equation}
N \equiv {\rm integer} \; + \frac{g^2}{8 \pi^2}\int_{\tau_0}^{\infty} 
	d\tau \int d^3 x E_i^a B_i^a \, ,
\label{Def_of_N}
\end{equation}
where we mean the lattice implementation of the integrals and $E$, $B$
as before.  In other words, we ``pre-cool'' the configuration for
cooling time $\tau_0$ and then measure $N_{\rm CS}$.  The pre-cooling is
intended to remove UV excitations without affecting the underlying IR
fields much.

Now consider $N$ in the abelian theory again.  The theory is linear, so
it is easy to analyze how cooling affects it.  A particular transverse
mode $A(k)$ evolves according to
\begin{equation}
\frac{dA(k)}{d\tau} = - \frac{\partial H}{\partial A(k)} = - k^2
	A(k) \quad \Rightarrow \quad A(k,\tau) 
	= e^{-k^2 \tau} A(k,0) \, .
\end{equation}
The propagator in Landau gauge becomes 
\begin{equation}
\langle A_i(k,\tau) A_j(l,\tau) \rangle = \left( \delta_{ij} - 
	\frac{k_i k_j}{k^2} \right) \frac{ e^{-2 k^2 \tau}}{k^2} 
	\delta(k+l) \, ,
\end{equation}
and the variance of $N$ is 
\begin{equation}
\langle N^2 \rangle = \frac{g^4 T^2 V}{64 \pi^4} \int \frac{d^3 p}
	{ (2 \pi)^3 } \frac{p^2 e^{-4 p^2 \tau}}{(p^2)^2} \, .
\end{equation}
So pre-cooling removes the UV noise from the definition of $N_{\rm CS}$.

\begin{figure}[t]
\centerline{\mbox{\psfig{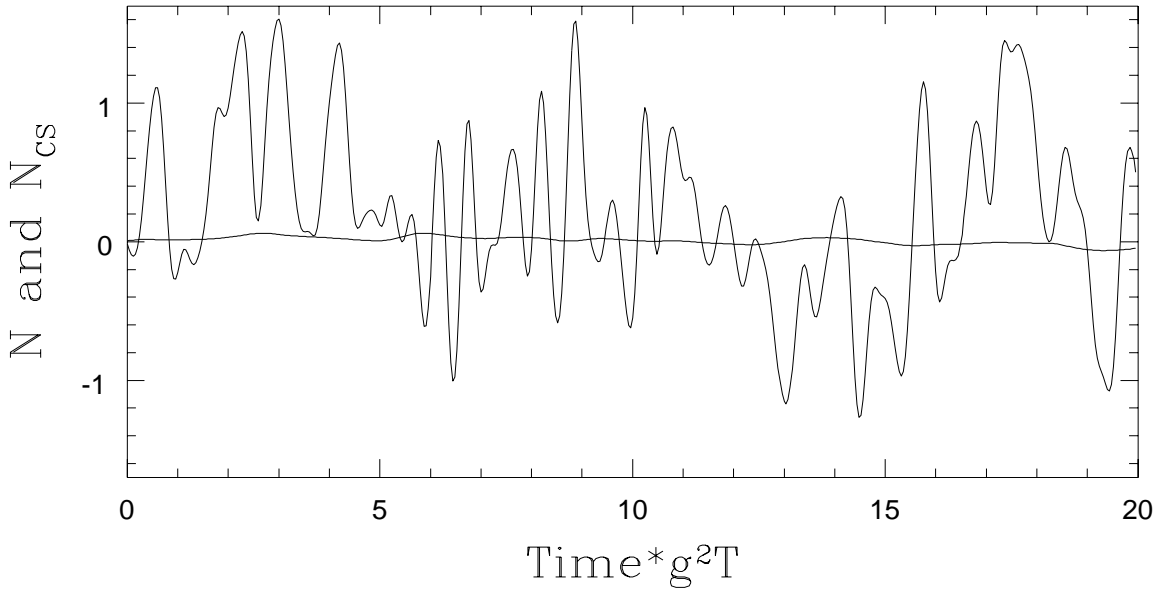}}}
\caption{ \label{NvsNCS} $N_{\rm CS}$ (wildly oscillating curve) and $N$
(curve which stays near 0) during a broken phase Hamiltonian trajectory,
in a $40^3$ box with $a=2/(5 g^2 T)$ (and using $\tau_0 = 3.2 / (g^2
T)^2$).  $N_{\rm CS}$ has a lot of UV noise, which is absent in $N$.}
\end{figure}

For comparison, Figure \ref{NvsNCS} shows $N$ and $N_{\rm CS}$ for the
same broken phase Hamiltonian trajectory; while $N_{\rm CS}$ varies
wildly on a short time scale due to UV fluctuations, $N$ is steady,
and shows that the infrared fields never stray far from the vacuum.

$N$ will not meet all the conditions we set out for $N_{\rm CS}$; for
instance it will violate condition 4. severely.  However, it will be
closer to a continuous function, since initial
cooling removes UV excitation from the configuration, leaving weaker and
more slowly varying fields for which the definition of $E_i^a B_i^a$ is
less problematical.  There is also less problem using coarsening with
this definition.  For instance, on a $28^3$ grid, single coarsening
after $\tau = (5/4) a^2$ and double coarsening after $\tau = 2.8
(2a)^2$, and using an $O(a^2)$ improved definition of $E^a_i B^a_i$, the
discontinuity in $N$ across the 
gradient flow separatrix is $0.9870 \pm 0.0029$
(drawing configurations from broken phase Yang-Mills Higgs theory at
$\beta_L = 7$, see next section).  To make $N$ a continuous function
modulo 1, the jump should have been 1.  By rescaling $N$ slightly, the
discontinuity is removed almost altogether.  The value of the
discontinuity for larger 
lattices and deeper cooling is even closer to 1 with less noise. 

\begin{figure}[t]
\centerline{\mbox{\psfig{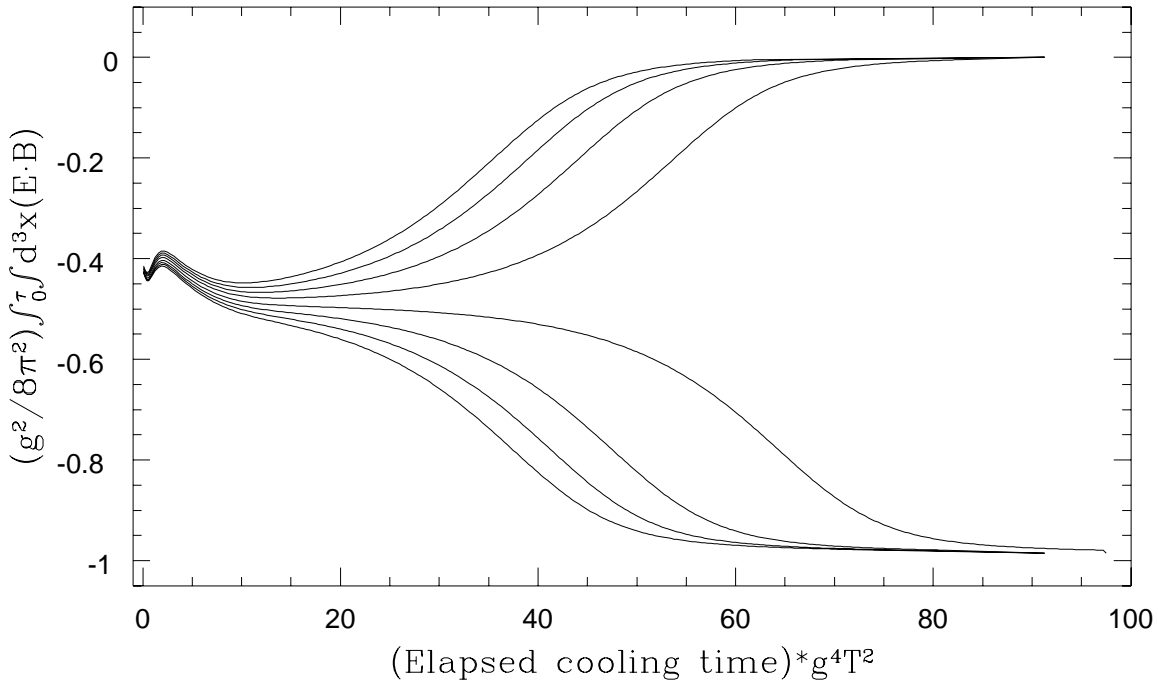}}}
\caption{Plot of $\int E\cdot B$ from $0$ to $\tau$, as a function of
$\tau$, for a series of points along a Hamiltonian trajectory as it
passes through the gradient flow separatrix. \label{asitcools}}
\end{figure}

Let us check that the new definition of $N$ will have $N \simeq 1/2$ on
the gradient flow separatrix.  A point right on the separatrix will
gradient flow to the saddlepoint 
configuration and stick there.\footnote{This saddlepoint is not the same
as Klinkhamer and Manton's sphaleron, because we are considering the
Yang-Mills Hamiltonian only, 
in a finite volume.  But we know such a saddle will exist by the
same argument Manton originally made for the existence of the sphaleron
\cite{Manton1}.}  Perturbing the starting configuration slightly off 
the separatrix, it
will cool to the saddle, miss slightly, and then slide off
to a vacuum.  The closer to the separatrix we start, the longer we will
stick in the saddle before we slide out.  There is some early $\tau$
(transient) contribution to 
$\int \int E_i^a B_i^a d^3 x d\tau$ while it is
approaching the saddle, and then there is a contribution, almost exactly
equal to $1/2$, as it rolls from the saddle to the vacuum.  By choosing
$\tau_0$ large enough, we miss the transient and pick up only the $1/2$,
and so the $N=1/2$ separatrix will correspond almost precisely with the
gradient flow separatrix.  
We illustrate this with data from a Yang-Mills theory simulation in
Fig. \ref{asitcools}.  The figure plots
\begin{equation}
\frac{g^2}{8 \pi^2} \int_0^{\tau} d\tau' \int d^3 x E_i^a B_i^a 
\end{equation}
(shifted so the vacuum will have integer $N_{\rm CS}$)
against $\tau$ for a series of points on a Hamiltonian trajectory as it
goes through the separatrix.  Each curve records the cooling process of 
a successive point on the Hamiltonian trajectory.  
We see that as the Hamiltonian trajectory 
approaches the (gradient) separatrix,
the cooling path stays near the saddlepoint for longer and longer.
When the Hamiltonian trajectory crosses the 
separatrix, the cooling paths
roll out of the saddle towards the other side.  The figure also shows
the early $\tau$ transient.  We want to choose a $\tau_0$ large enough to
eliminate this transient, so $N \simeq 1/2$ will hold for a configuration
which starts almost exactly on the gradient flow separatrix.  

We should make sure that the result for $\Gamma_d$ will be
independent of $\tau_0$, once $\tau_0$ is large enough to kill the UV
problems.  We remind the reader that $\Gamma_d$ is computed by choosing
an $\epsilon \ll 1$, and computing the probability to be within
$\epsilon/2$ of 1/2,
\begin{equation}
P_\epsilon = \int_{(1-\epsilon)/2}^{(1+\epsilon)/2} P_N(x) dx \, .
\end{equation}
Also, one computes the time rate of change of $N$, $\langle | dN/dt |
\rangle$, evaluated for configurations with $N \simeq 1/2$.  The rate
$\Gamma_d$ is then
\begin{equation}
\Gamma_d = \frac{1}{V} \frac{P_\epsilon}{\epsilon} \langle | dN/dt |
	\rangle \, ,
\end{equation}
times the dynamical prefactor.

Increasing the cooling time will decrease $P_\epsilon$.  The reason is
that configurations which are near the saddlepoint at cooling
time $\tau_0$ are spreading out from each other as cooling time
progresses, as figure \ref{asitcools} illustrates.  The rate of the
spreading is given by the unstable frequency squared 
$(\omega_-)^2$ of the saddle\footnote{which does not equal 
$(\omega_-)^2$ of the Klinkhamer Manton sphaleron}.  
Increasing $\tau_0$ makes them spread
further before we measure their $N_{\rm CS}$; so the sample of states
will be more diluted, by a factor of $\exp(- \Delta \tau_0 / 
(\omega_-)^2)$.  This reduces $P_\epsilon$ by the same factor.  However,
we measure $dN/dt$ by choosing two neighboring configurations on a
Hamiltonian trajectory and finding the difference in their values of
$N$.  The spread between these at cooling time $\tau_0$ will also
increase as we increase $\tau_0$, by the same amount; so $\langle |
dN/dt | \rangle$ will go up by $\exp( \Delta \tau_0 / (\omega_-)^2)$.
Hence, $\Gamma_d$ will be $\tau_0$ independent.  This means, however,
that neither $\langle | dN/dt | \rangle $ nor $P_N(x)$ have simple
physical interpretations.  

There is a modification of the above reasoning if $\tau_0$ is too short
to eliminate the early transient; namely, a contribution to $dN/dt$ due
to the time evolution of the transient.  In the complete calculation
this will be compensated for because the dynamical prefactor will differ
from the gradient flow value by a $\tau_0$ dependent amount, which
becomes non-negligible at the same time the transient contribution to
$dN/dt$ does.  In a complete calculation, $\Gamma_d$ will be independent
of $\tau_0$, as we argued in the last section.

We end this section by discussing briefly why we choose to define $N$
based only on the Yang-Mills fields and using the Yang-Mills Hamiltonian 
for the cooling, rather than including the Higgs field.
Doing so is reasonable because $N_{\rm CS}$ should
be defined as a function of the gauge fields alone.  Also, cooling all
the fields under the full Hamiltonian is problematic, because the UV
fluctuations of the gauge and Higgs fields renormalize the Higgs mass
squared \cite{KRSearly}.  The bare potential needs a large, negative
mass squared counterterm.  However, the UV fluctuations are the first
casualty of the cooling process, and so they stop generating a thermal
Higgs mass squared, early in the cooling.  To keep the minimum of the
Higgs potential from changing radically, one would need to vary the bare
Higgs potential in a complicated way as the cooling progressed.  Other
important fluctuation induced effects are also lost; for instance, the
cubic term which makes the phase transition first order disappears as we
cool the excitations.  Depending on how we handle the Higgs potential
during cooling, we will either cause symmetry to break during cooling
when we start in the symmetric phase, or cause it to be restored when we
start in the broken phase; we cannot avoid both, because there is a
range of temperatures where each phase is metastable.  Cooling
only the gauge fields avoids this complication.

\section{Monte-Carlo calculations}
\label{Montesec}

Here we discuss details of performing the calculation of $P_N(x)$ and
$\langle | dN/dt | \rangle $ using the definition of $N$ from
Eq. (\ref{Def_of_N}).  We also discuss the computation of the dynamical
prefactor, with and without ``adding'' hard thermal loops.
Readers who don't do this kind of calculation will want to skip all 
but the first subsection and go to the results.

\subsection{multicanonical Monte-Carlo:  the idea}

We need to compute the probability distribution $P_N$ of Chern-Simons
number (really, the Chern-Simons number of a precooled configuration,
defined in Eq. (\ref{Def_of_N})).  
The probability density that $N = x$ is
\begin{equation}
P_{\rm N}(x) = \lim_{\delta x \rightarrow 0} \frac{1}{Z \delta x} \int
	{\cal D} U {\cal D} u {\cal D} \Phi e^{- \beta H(U,u,\Phi)} 
	\Theta(N(U) - x) \Theta(x + \delta x - N(U)) \, ,
\label{canonMC}
\end{equation}
where $U,u,\Phi$ are the SU(2) connection, the U(1) connection, and the
Higgs field, and $N(U)$ is defined in Eq. (\ref{Def_of_N}).  Here 
$\int {\cal D} U {\cal D} u {\cal D} \Phi$ means the integral over the
value of each field at each lattice site, and $Z$ is the value of the
integral without the step functions.  On an $N^3$ lattice, this is
a $16 N^3$ dimensional integral, which for $N=40$ is $1024000$
dimensions.  For this reason we turn to Monte-Carlo integration.  In a
canonical Monte-Carlo integration we generate a sample of
configurations drawn with weight 
\begin{equation}
e^{- \beta H(U,u,\Phi)} {\cal D} U {\cal D} u {\cal D} \Phi \, ,
\end{equation}
 and replace the integral with a
sum over that sample.  This will not do in the present context,
because we want to know $P_{\rm N}(x)$ even where it is exponentially
small.  To get a good sampling there would require generating an
exponentially large sample.

We evade this problem
by doing a multicanonical Monte-Carlo calculation\cite{MCMC}.  
We rewrite Eq. (\ref{canonMC}) as
\begin{eqnarray}
P_{\rm N}(x) & = & \lim_{\delta x \rightarrow 0} \frac{1}{Z \delta x} 
	\int \left( {\cal D} U {\cal D} u {\cal D} \Phi 
	e^{- \beta H(U,u,\Phi)} e^{f(N(U))} \right) \times \nonumber \\
& & \qquad \qquad \left( e^{-f(N(U))} \Theta(N(U) - x) 
	\Theta(x + \delta x - N(U)) \right) \, ,
\label{MCMC}
\end{eqnarray}
with $f(x)$ some function we are free to choose.
Now we generate a sample of configurations drawn with weight 
\begin{equation}
e^{- \beta H(U,u,\Phi)} e^{f(N(U))} {\cal D} U 
	{\cal D} u {\cal D} \Phi \, , 
\end{equation}
and replace the integration in 
Eq. (\ref{MCMC}) with a sum over this sample, with the term in the
second set of parenthesis as the argument of the sum.  By choosing $f(x)
\simeq - \ln P_{\rm N}(x)$, we push the exponential suppression from the
sampling into the integrand.  The quality of the integration is now
limited by how quickly we can generate a quality sample with this
weight, and how well we can choose $f(x)$, which we must do by some form
of bootstrapping.

A usual way to generate a canonical ensemble is by some Markov process.
Given a configuration $C_1$ and a realization $\xi$ of some random noise
distribution, the process returns a new configuration $C_2 = M(C_1,
\xi)$.  Define the probability to return a particular $C_2$ as
\begin{equation}
{\cal P}(C_1 , C_2) \equiv \int d\xi \: \delta( M(C_1 , \xi) , C_2 ) \, .
\end{equation}
If $M$ satisfies detailed balance,
\begin{equation}
\frac{ {\cal P}(C_1,C_2)}{ {\cal P}(C_2, C_1)} = \exp( \beta (
	H(C_1) - H(C_2))) \, ,
\end{equation}
then iterating the Markov process generates the canonical distribution.
We get the best statistics if applying $M$ is numerically cheap and if
the return value differs as much as possible from the starting value.

If we have a Markov process which generates the canonical ensemble, we
make it generate the multicanonical ensemble with weight function $f(C)$
by making the following modification.  Given a configuration $C_1$,
generate $C_2 = M(C_1, \xi)$.  Accept $C_2$ as the next configuration in
the sequence with probability
\begin{equation}
{\rm min}(1, \exp(f(C_2) - f(C_1))) 
\end{equation}
and otherwise reject it and make $C_1$ the next configuration.  This
changes the detailed balance relation for the sequence to incorporate
$f$ into the weight.  We may no longer want $M(C_1,\xi)$ to differ from
$C_1$ by as much as possible, though, because that may make the reject 
rate very large.  Instead we want $| f(M(C_1 , \xi)) - f(C_1) | \sim 1$.

\subsection{note on algorithm}
\label{algorithm_sec}

We need to do two kinds of things.  The first is to evaluate the path
integral for dimensionally reduced 3-D Yang-Mills Higgs theory.
The second is to study the dynamics of the 3+1 dimensional, classical
theory.  We comment briefly on the connection between the two; in
particular we should compare the partition function of the classical
theory to the path integral for the dimensionally reduced theory, a
comparison first made by Ambj{\o}rn and Krasnitz \cite{AmbKras}.

The partition function of
classical, 3+1 dimensional Yang-Mills Higgs theory (not worrying about
the difference between lattice and continuum, which will not be
important here) looks like
\begin{eqnarray}
\label{Partitionfunc}
Z & = & \int {\cal D} A_i {\cal D} \Phi {\cal D} E_i {\cal D} \Pi
	\delta ( (D_i E_i)^a + (g/2)( i \Pi^\dagger \tau^a \Phi + h.c. ) 
	 ) \exp(-H(A,\Phi,E,\Pi)/T) \, , \\
H & = & \int d^3 x \left( \frac{B^2}{2} + \frac{E^2}{2} + (D_i \Phi)^2 
	+ \Pi^2 \right) \, .
\end{eqnarray}
where $E$, the electric field, is the conjugate momentum of $A$, and 
$\Pi$ is the conjugate momentum of $\Phi$.  The delta function enforces
Gauss' Law.  We can implement Gauss' Law by means of an adjoint valued 
Lagrange multiplier $A_0^a$, giving \cite{AmbKras}
\begin{equation}
\label{Partition_w_A0}
Z = \int {\cal D} A_i {\cal D} \Phi {\cal D} E_i {\cal D} \Pi
	{\cal D} A_0
	\exp(i A_0^a ((D_i E_i)^a + (g/2) \Pi^\dagger \tau^a \Phi 
	+ h.c.)/T) \exp(-H(A,\Phi,E,\Pi)) \, .
\end{equation}
The integrals over $E$ and $\Pi$ are rendered Gaussian; doing them gives 
\begin{eqnarray}
\label{Partition_no_E}
Z & = & \int {\cal D} A_i {\cal D} \Phi {\cal D} A_0 
	\exp(-H(A,\Phi,A_0)/T) \, , \\
H & = & \int d^3 x \left( \frac{B^2}{2} + \frac{(D_i A_0)^2}{2}
	+ (D_i \Phi)^2 + \frac{g^2}{4} A_0^2 \Phi^2 \right) \, .
\end{eqnarray}
So the thermodynamics of the 
classical theory with Gauss' Law is governed by the 3-D dimensionally
reduced path integral, but including the $A_0$ field at zero (bare)
Debye mass.  We could get the path integral without the $A_0$ field
if we did not enforce Gauss' Law.  There are two
choices; either we treat the dynamics dropping Gauss' Law, or we include
the $A_0$ field in the 3-D Monte-Carlo parts of the calculation.

The conservative approach is to include Gauss' Law in the dynamics,
which means including an $A_0$ field in the thermodynamics.  The easiest
canonical Monte-Carlo Markov process in this case is a short Hamiltonian
trajectory starting with randomized but constraint respecting $E$ and 
$\Pi$ fields, with the multicanonical accept reject steps 
inserted between evolutions.  This is a ``constrained molecular
dynamics'' algorithm, and the problem of drawing $E$ and $\Pi$ from the
thermal distribution respecting the constraints is addressed in 
\cite{Moore1}.  Alternately we could use heat bath and
overrelaxation updates on the system described in
Eq. (\ref{Partition_no_E}).  Neither approach is terribly efficient.

The other option is to assume that Gauss' Law is not very important to
the dynamics, and not enforce it when we draw momenta $E$ and $\Pi$ from
the thermal ensemble.  This assumption is certainly justified with
regards to measuring $\langle |dN/dt| \rangle$.  What the combination of
$E$ and $\Pi$ forced zero by Gauss' Law would do if we did not set them
zero is to gauge rotate the fields, but not the momenta.  On long time scales
that might be important, but at leading order in the length of a short 
Hamiltonian trajectory it only changes the gauge of the final
configuration.  Since $N$ is a gauge invariant object, this does not
matter.  A less rigorous but more physically intuitive way to see the
unimportance of Gauss' Law to $\langle |dN/dt| \rangle$ is to note that
it is roughly the instantaneous value of $\int d^3 x E_i^a B_i^a$.  Now
the magnetic field is transverse by the Bianchi identity, so only the
transverse components of the electric field contribute to $dN/dt$.  But
Gauss' Law only depends on the longitudinal components, so $dN/dt$ should
be the same whether or not we enforce it.

It is less clear whether Gauss' Law will have a
role in setting the dynamical prefactor, since it depends on longer time
dynamics; but we are getting the dynamical prefactor 
wrong anyway if we don't enlarge the
system somehow to account for hard thermal loop effects properly.  We
should deal with these two questions together.

The chief advantage of not enforcing Gauss' Law is that there are very
efficient update algorithms for the path integral without either the $E,
\Pi$ or $A_0$ fields, for instance the one developed by Rummukainen
et. al. \cite{KLRSresults}.  We adopt their lattice action, which is
the standard Wilson 3-D Yang-Mills Higgs action, but we add 
a noncompact U(1) field.  We use their update, which must be extended 
to include the noncompact U(1) field.  In the noncompact formulation, 
the U(1) gauge field on a link is represented by a single real
number $B_i(x)$, and the terms in the action which depend on the U(1)
field are
\begin{eqnarray}
H & \supset & \frac{a}{2g^2 z} \sum_x \left[ \sum_{i>j} \left(
	B_i(x) + B_j(x+i) - B_j(x) - B_i(x+j) \right)^2 \right]
	\nonumber \\ & &
	- a \sum_x \sum_i \phi^\dagger(x) U_i(x) \exp(ia B_i(x)) 
	\phi(x+i) \, ,
\end{eqnarray}
where $a$ is the lattice spacing, 
$z = \tan^2 \Theta_W$, and $U_i(x)$ is the SU(2) connection on the
$i,x$ link.  The compact formalism is the same except that
$(1/2)(\sum B)^2$ is replaced by $a^{-2}(1 -\cos(a\sum B))$.  
In the compact case the energy, as a function of one $B$, is a 
trigonometric function.  It is easy to
perform an exact heat bath or overrelaxation step \cite{KLRSSU2U1}.  
For the noncompact case the energy is the sum of a quadratic and a
trigonometric function and it is not easy to perform an exact overrelaxation
or heat bath update.  We perform the update based just on the (much
larger) quadratic term and include the trigonometric term by
an accept reject step.  The accept rate is quite high, so the cost to
algorithm efficiency is low.  We also occasionally gauge transform to
bring the $B$ fields
towards Coulomb gauge so that the typical $B$ is close to
zero and the series expansion of $\exp(ia B_i(x))$ converges quickly.

We always apply $O(a)$ improvement to the lattice action as described in
\cite{Oapaper,Oa2}.  Whenever we refer to physical units in this paper
they are always related to the lattice ones through $O(a)$ improved
relations.  The improvement is essential to achieving the numerical
accuracy we want at reasonable lattice spacings, and it makes small
spacing extrapolations of most quantities unnecessary.  For instance, we
have computed the jump in $\phi^2$ at the equilibrium temperature for $a
= 4/7g^2 T$ ($\beta=7$) and $a=2/5g^2 T$ ($\beta=10$) to see how
sensitive results are to varying $a$.  The results are  
$ ( \Delta \phi^2  / g^2 T^2 ) =
2.51 \pm .02$ and $2.56 \pm .03$ respectively; the ($O(a^2)$) errors are
smaller than the statistical errors we will be able to achieve for
$\Gamma_d$, so lattice spacing errors are under control.

We have also used the technique of \cite{particles} to study the
influence of hard thermal loops on the dynamical prefactor.  The idea is
to add a large number of weakly interacting, ballistic charged particles
to the lattice system, which reproduce the effects of the hard degrees
of freedom left out when we set up our lattice.  We refer the reader to
\cite{particles} for details.  This approach demands that we apply
Gauss' Law, and we must use the (Hamiltonian) thermalization algorithm
as our Markov process.  This is very inefficient, so we can only use
the technique to study the dynamical prefactor, not the flux through the
separatrix.  

\subsection{finding $T_c$}

At what temperature should we study the sphaleron rate in the broken
phase?  We choose the equilibrium temperature for the phase transition.
This is appropriate if the latent heat liberated during the cosmological
electroweak phase transition is sufficient to reheat the universe to
$T_c$, which would be the case if the latent heat were large or the
supercooling were small.  It is in fact not clear whether this will be
the case, in the MSM or in more viable extensions.  
We will choose $T_c$ anyway, because 
we then have a well specified question with only
one free parameter $\lambda / g^2$; besides, complete reheating may
generically occur in supersymmetric extensions with a light stop.  (It
also may not; see \cite{Cline_inprogress}.)

In the effective 3-D theory we
actually find the critical Higgs mass squared, which is related through
the dimensional reduction procedure to the critical temperature.
Since this involves comparing the thermodynamical 
favorability of the broken and
symmetric phases, it necessarily involves some multicanonical
technique.  We will use a particularly simple approach, somewhat similar to
the one used in \cite{Fodor}.  It is based on the fact that, in a very
long rectangular box, values of $\int \Phi^\dagger \Phi / V \equiv
\langle \phi^2 \rangle$ intermediate between the symmetric and broken
phase values are obtained by having a mixed phase configuration, where
part of the volume is in the symmetric phase and the rest is in the
broken phase.  

The free energy, as a function of $\langle \phi^2
\rangle$, is $- T \ln(P)$, with $P$ the probability to have that value  
of $\langle \phi^2 \rangle$.  In the range of intermediate $\langle
\phi^2 \rangle$, the free energy will vary linearly with $\langle \phi^2
\rangle$, since a change of $\langle \phi^2 \rangle$ represents a change
of how much bulk free energy comes from one phase and how much comes
from the other phase.  The slope of the linear relation tells the free
energy difference between the two bulk phases.  This linear regime
breaks down where $\langle \phi^2 \rangle$ comes close to the value in
one or the other phase, since the phase interfaces then get close enough
together to interact.

Our approach is to add a $\langle \phi^2 \rangle$ dependent contribution
to the Higgs mass, which we achieve by adding to the Hamiltonian a new
nonlocal term, 
\begin{equation}
\eta N^3 ( \frac{1}{2 N^3} \sum \Phi^\dagger \Phi )^2 \, .
\label{etaterm}
\end{equation}
Choosing $\eta$ to be positive means that, if most of space is in the
broken phase, the Higgs mass is heavier and the symmetric phase is
favorable, whereas if most of space is in the symmetric phase, then the
Higgs mass is smaller and the broken phase is favorable.  The free
energy is then a quadratic function, and the effective Higgs mass
squared at its minimum, including the contribution from the $\eta$ term,
gives the equilibrium Higgs mass.  The added term is simple enough that
we can modify the canonical update to incorporate it, and Monte-Carlo
evolution will then naturally settle in a mixed phase configuration
whose value of $\langle \phi^2 \rangle$ tells us the equilibrium Higgs
mass, 
\begin{equation}
m_{\rm H}^2({\rm equilibrium}) = 
	m_{\rm H}^2 ( \langle \phi^2 \rangle = 0 ) + \eta 
	\langle \phi^2 \rangle  \, .
\end{equation}
This approach can be viewed as a type of multicanonical Monte-Carlo with
the $\eta$ term as the multicanonical reweighting.

In practice we start with a long but very narrow box and a high value of
$\eta$, to get a preliminary value.  The narrowness is necessary to make
it easy to nucleate a bubble of one phase in the other.  To get the
large volume limit, though, we need to go to a wider box; we necessarily
need results in a regime where one phase cannot easily nucleate in the
other.  We get an
initial condition for a box an integer number of times wider in each
short direction by extending the final configuration in the skinny box
periodically.  We also use a smaller value of $\eta$ to improve the
resolution of the determined $m_{\rm H}^2$.  
Our final values for $m_{\rm H}^2$ are
typically taken with a box of dimension, in physical units, of $(16/g^2
T)^2 \times (96/g^2 T)$, easily large enough to achieve the large volume
regime.  

\subsection{practicing with the symmetric phase in small volume}

Before presenting the determination of the broken phase sphaleron rate,
we will do a ``practice run'' on a system where we can get good
statistics more quickly, which is the symmetric phase in a small enough
volume to suppress topology change.  This problem is almost certainly of
no phenomenological significance, but it will let us test the $\tau_0$
dependence of our technique and to study whether hard thermal loops do
indeed modify the dynamical prefactor.

For the time being we drop the U(1) subgroup.  We choose a very weak 
scalar self-coupling of $x=0.025$, and a large Higgs mass squared 
($m_{\rm H}^2 \simeq 1.5 g^4 T^2$), so we will be firmly in the symmetric 
phase.  We use a $12^3$ lattice with a lattice spacing
of $a = 1/(4 g^2 T)$ ($\beta_g = 16$), so the physical volume 
is $(3/g^2 T)^3$.  This volume is small enough that topology change is
suppressed and the broken phase techniques we are developing are
applicable, but not so small that it will be hard to gather good
statistics.  It is also small enough that numerical costs are not
overburdening, so we will enforce Gauss' Law (and use the less efficient
update algorithm).

We measure $N$ by cooling for $\tau = (5/4) a^2$, coarsening once, and
using an $O(a^2)$ improved definition of $E_i^a B_i^a$ during the
subsequent cooling.  The first thing we do is to determine the actual
discontinuity in $N$ across the gradient flow separatrix, which will
differ from 1 because of lattice artifacts.  To do this we generate an 
ensemble of configurations on the gradient flow separatrix.  For each, we
perform a short Hamiltonian evolution which crosses the separatrix, and
we measure $N$ at closely spaced intervals during the crossing to determine
the discontinuity; it is $\Delta N =
0.982 \pm 0.005,$ where the error is the standard deviation\footnote{It
is important to compare the extrapolation based on a fit of the last few
points on one side of the discontinuity with the value on the other, to
remove errors from the time step size of the Hamiltonian evolution used
to find the discontinuity.}.  This tells us how to rescale
all further measurements of $N$ so that there will be an
integer discontinuity at the separatrix (up to acceptably small noise).  

\begin{figure}[t]
\centerline{\mbox{\psfig{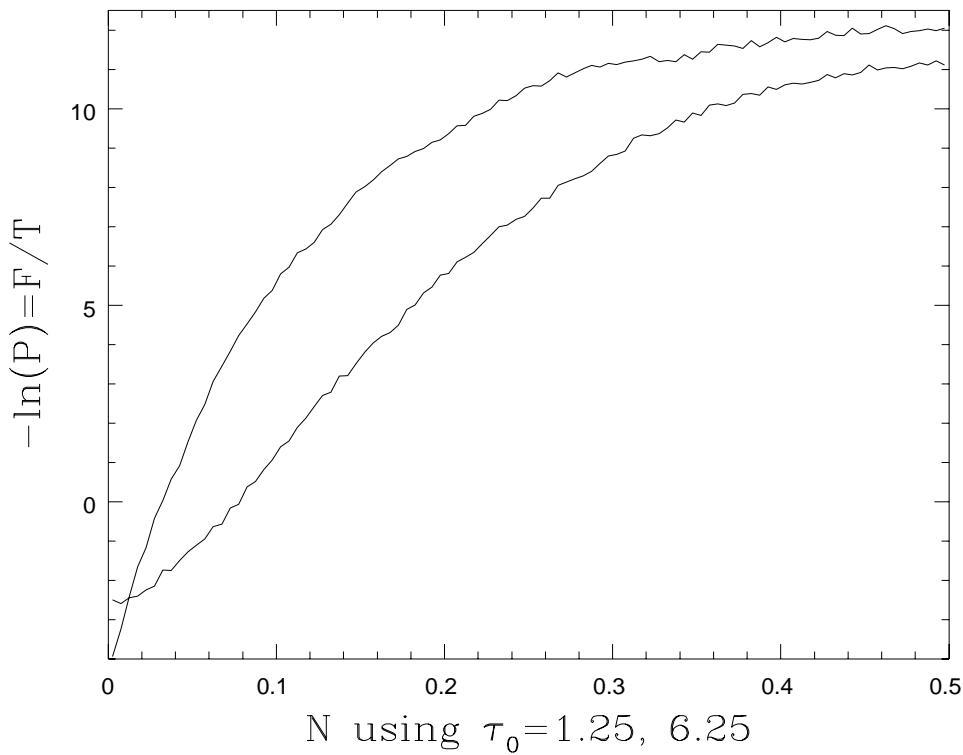}}}
\caption{\label{compare_tau0} Probability distribution of $N$ for two
values of $\tau_0$:  $\tau_0 = 1.25 a^2$ (the rounder curve which is
lower at $N=1/2$); and $\tau_0 = 6.25 a^2$ (the curve which is higher at
$N=1/2$).}
\end{figure}

Next, we measure the probability distribution for $N$.  We only need to
do this in the range $0 \leq N \leq 1/2$, by periodicity; so we add $1$
to all negative values of $N$ and then put $N>1/2$ into range by $N
\rightarrow 1 - N$.  We use a continuous, piecewise linear reweighting
function, with the widths of the linear pieces chosen by hand and
the slopes of each segment determined by an automated bootstrapping 
procedure.  We computed the probability distribution for two values of
$\tau_0$; $\tau_0 = 1.25 a^2$ and $\tau_0 = 6.25 a^2$.  For the
$\tau_0 = 1.25$ data we also recorded $N ( \tau_0 = 6.25 ) $; 
the data can therefore be used to get the probability distribution for 
either choice of $\tau_0$.

The probability distributions are compared in Figure
\ref{compare_tau0}.  The distributions are clearly different.  Cooling
longer before integrating $E_i^a B_i^a$ concentrates probability around
$N=0$ and thins out the large $N$ configurations.  It is less likely to
have $N$ within some range of $1/2$ for the 
larger $\tau_0$.  But this does not mean that the
different $\tau_0$ values give different values for the diffusion
constant, as we still have not included $\langle | dN/dt | \rangle$ or
the dynamical prefactor.  

To get the dynamical prefactor and $\langle |dN/dt| \rangle$ we first
need a sample of configurations very close to $N=1/2$.  We get them by
multicanonically sampling, not necessarily with the same reweighting
function used to find the probability distribution of $N$.  Then, for
each, we choose momenta out of the appropriate distribution and perform
a Hamiltonian evolution, with the algorithm of \cite{Ambjornetal}.  Once the
Hamiltonian trajectory has settled into the neighborhood of a vacuum, we
return to the configuration before we started the Hamiltonian evolution
and reverse the sign of the momentum; then we evolve.  Since momenta are
odd and fields are even under time reversal, this computes the
Hamiltonian trajectory in the backwards time direction.
We determine $\langle | dN/dt | \rangle$ from the first
time step in each time direction, and the dynamical prefactor from the
number of $N=1/2$ crossings, as discussed in the previous section.
It is also easy to find how many times the Hamiltonian trajectory
crosses the gradient flow separatrix, since $\int E_i^a B_i^a d^3 x 
d\tau$ abruptly
changes sign when it does; so we can also identify how well correlated
crossings of the $N=1/2$ and gradient flow separatrices are.  (We
cannot directly determine the gradient flow separatrix prefactor if the
correlation is not good, because our sampling procedure is for the
$N=1/2$ separatrix and not the gradient flow one.)

\begin{figure}[t]
\centerline{\mbox{\psfig{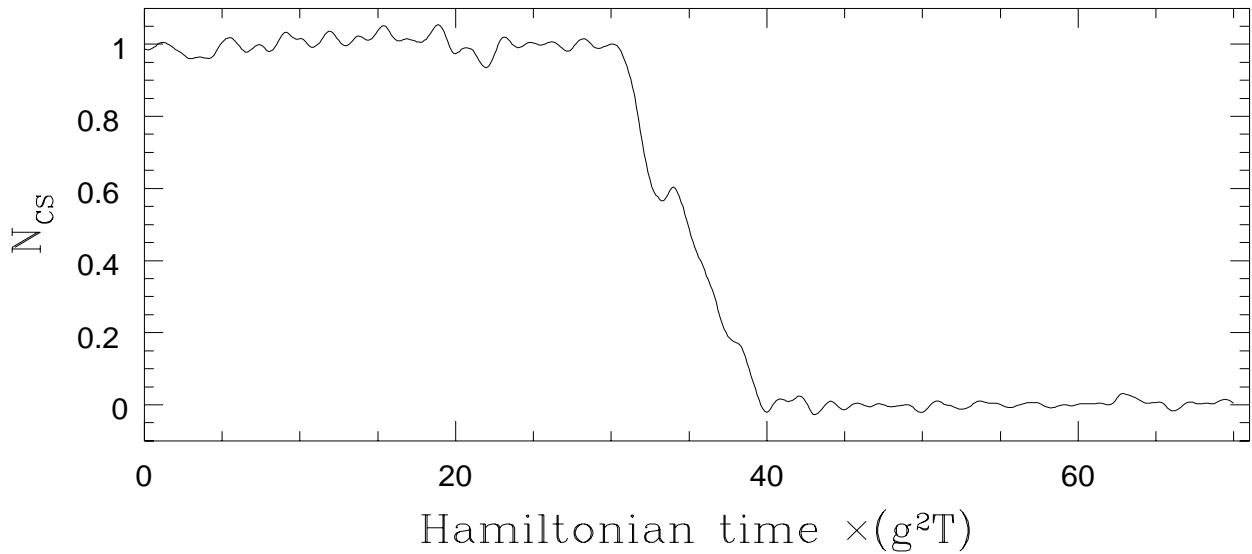}}}
\caption{\label{Hamtrajectory}
A winding number changing section of a Hamiltonian trajectory in
constrained volume Yang-Mills theory.  After the crossing the system
settles immediately into the neighborhood of a topological vacuum.}
\end{figure}

Figure \ref{Hamtrajectory} gives an example of a Hamiltonian trajectory
developed in this way, with $\tau_0 = 3.75 a^2$.  It is clear in this
figure that once the trajectory settles in the neighborhood of a
topological vacuum, in the sense that $N \sim 0$, then it stays there
for some time.  Since the maximal Lyapunov exponent of the Yang-Mills
Higgs system is known to be about $0.3 g^2 T$
\cite{MullerTrayanov,Biro,AmbKras}, the direction of the next permanent
change in $N_{\rm CS}$ will surely be statistically independent from the
previous one.  This fact is essential to the whole approach; our most
basic assumption is
that the very long time $N_{\rm CS}$ diffusion is made up of a
series of statistically independent steps, and we only need find out how
frequently one of those steps is taken (which is what we are computing
when we say we are computing $\gamma_d$).

Also note from Figure \ref{Hamtrajectory} that there is no
``overshoot'' after reaching $N=0$, no sign that the trajectory is
continuing in the direction of the next separatrix.  This is true of all
trajectories we have studied, both in finite volume and in the broken
electroweak phase, which confirms the absence of prompt multiple
crossings.

\begin{table}
\begin{tabular}{|c|c|c|} \hline \hline
quantity & value at $\tau_0=1.25a^2$ & 
	value at $\tau_0=6.25a^2$ \\  \hline
$\ln P(|N-0.5|<.05)$ & $-13.39 \pm 0.22$ & 
	$-14.30 \pm 0.18$ \\ \hline
$|dN/dt|$ & $(.121 \pm .004)g^2 T$ & 
	$(.174 \pm .005)g^2 T$ \\ \hline
prefactor & $.52 \pm .03$ & 
	$.554 \pm .024$ \\ \hline
$\ln(\Gamma/(\alpha_w T)^4)$ & $-7.03 \pm .23$ &
	$-7.51 \pm .19$ \\ \hline
$\Gamma$ & $(.0009 \pm  .0002) \alpha_w^4 T^4$ & 
	$(.0005 \pm .0001) \alpha_w^4 T^4$ \\ \hline
\end{tabular}
\caption{\label{YMtable} Ingredients and results for the $N_{\rm CS}$
diffusion rate in a cubic, periodic volume $3/g^2 T$ on a side.  The
measurements with two values of $\tau_0$ disagree at $1.6 \sigma$.}
\end{table}

Our final results for this small volume system are presented 
in Table \ref{YMtable}.  Note that both the probability to be near
the separatrix and $\langle |dN/dt| \rangle$ vary quite a bit when we
change $\tau_0$, but in opposite directions.  Also, the shorter cooling 
leads to a smaller dynamical prefactor, though for the volume and
cooling considered here the difference does not turn out to be
large.  The determined value of $\Gamma$ for the two values of $\tau_0$
differ at $1.6 \sigma$.  This does not bother the author since
it is the first statistical fluctuation above $1.5 \sigma$ he has
encountered since starting numerical work; he had one coming.  

We can
make a more stringent check of the $\tau_0$ independence of the method
by using the data set taken with $\tau_0=1.25 a^2$ to determine the
probability distribution of $N$ 
for each value of $\tau_0$, since we recorded
$N(\tau_0 = 6.25 a^2)$ at each point while developing this data set.
This data set gives a slightly different probability to be at large
$N(\tau = 6.25 a^2)$; the $-14.30$ in the table becomes $-13.88$.  Using
this number, we get $\ln(\Gamma/(\alpha_w T)^4)=-7.09$.  To compare to
the $\tau_0=1.25 a^2$ data we must remember that the probability
distributions are now $100\%$ correlated, so the expected difference is
just from statistical errors in $\langle | dN/dt | \rangle$ and the
prefactor.  The difference in the logs of the rates is $.06 \pm .09$, so
they do agree within error.  
The results for $\Gamma_d$ are indeed $\tau_0$ independent.

How do hard thermal loops change the rate?  At the level of
thermodynamics, hard thermal loops become just a Debye mass.  They make
the $A_0$ field heavier, which pushes it further to the regime where it
decouples.  So their thermodynamic influence is very small.  Hence, they
will barely change the probability distribution of $N$, which will have
a well behaved large HTL strength limit.  Similarly, as we have
discussed, $\langle | dN/dt | \rangle$ does not depend on longitudinal 
$E$ fields, and hence depends on the Debye mass only through its
thermodynamic influence on the gauge fields.
The flux through the separatrix should depend weakly on hard
thermal loops and should have a good large HTL limit.

The dynamical prefactor is a completely different matter.  It is a
dynamical quantity which depends on unequal time correlators,
potentially over quite long times.  Hard thermal loops will change the
time evolution of infrared degrees of freedom on all time scales longer
than the inverse plasma frequency.  Arnold, Son, and Yaffe (ASY) have argued
\cite{ArnoldYaffe} that hard thermal loops will suppress the baryon
number violation rate by a factor parametrically of order 
$(g^4 T^2 / m_D^2)$,
because the number of crossings of the separatrix per permanent $N_{\rm
CS}$ change will be of order $m_D^2/g^4 T^2$.  That is, they predict that
turning on hard thermal loops will reduce the dynamical prefactor by an
amount linear in $m_D^2$ when it is large.
Their arguments have recently been verified in the symmetric phase
\cite{particles}.  Now we need to check what the hard thermal loops do
in the broken phase.  

\begin{figure}[t]
\centerline{\psfig{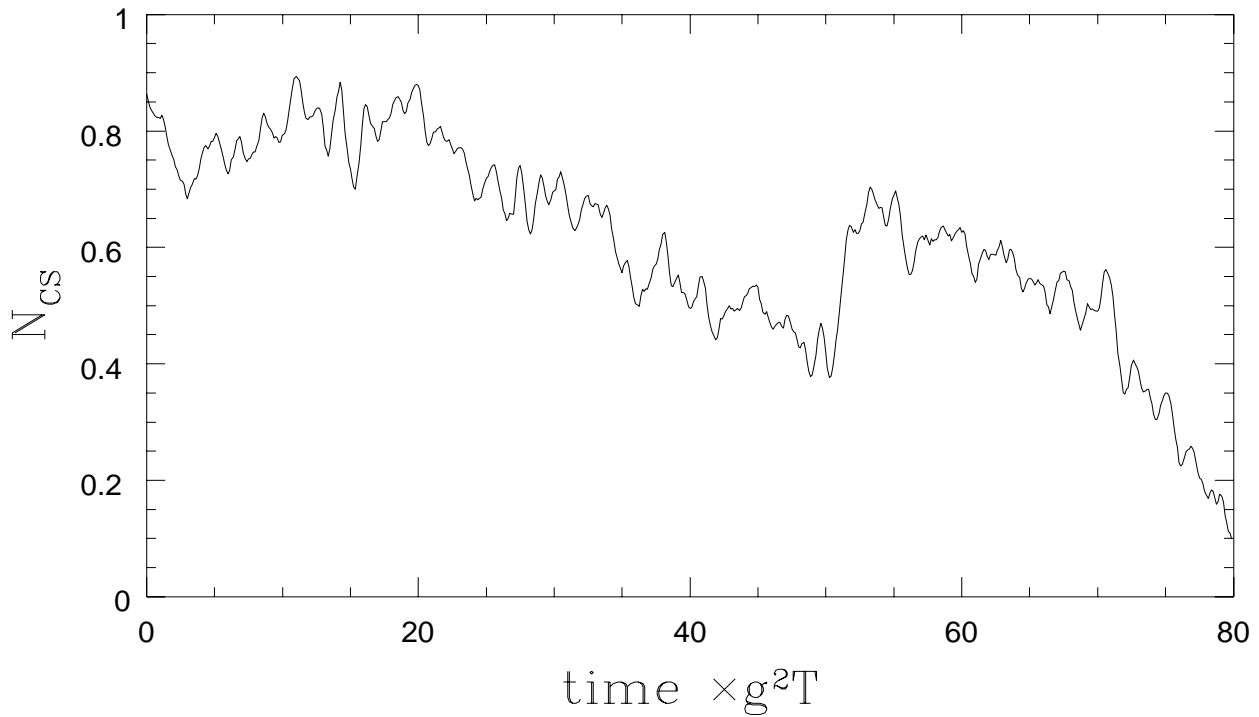}}
\caption{\label{HamwithHTL} Hamiltonian trajectory through the
separatrix in the symmetric phase in small volume, including hard
thermal loops with $m_D^2 \sim 43 g^4 T^2$.  Plasma oscillations are
clear, and there are numerous crossings of the separatrix.}
\end{figure}

To address this question we include the hard thermal loops by using the
technique developed in \cite{particles}.  We add a large number of new
``particle'' degrees of freedom to the system, which propagate the hard
thermal loop effects.  For now we will be happy to know what the hard
thermal loops do when they are extremely strong, so we can understand
the parametric limit in which the ASY arguments should hold.  With this
in mind we put in particles of charge $Q=0.1$ and number density
$50/a^3$,\footnote{See \protect{\cite{particles}} for the implementation
and the definitions of these quantities, and their relation to the Debye
mass.}
which give a Debye mass of $43 g^4 T^2$, an enormous number about 10
times the value in the MSM at a realistic value of $g^2$.  A Hamiltonian
trajectory crossing the separatrix is shown in Fig. \ref{HamwithHTL}.
The qualitative features are indeed the same as Arnold, Son, and Yaffe
predict, see Fig.8 of \cite{ArnoldYaffe}; plasma oscillations drive the
system across the separatrix numerous times.  The dynamical prefactor is
correspondingly significantly smaller than without hard thermal loops;
for these parameters it is about $0.16$.  
Note however that we have had to add
truly huge HTL effects to achieve this value, so at the realistic value
the effect may not be too significant.  We will study this question for
the broken phase case in the next section.

How do our nonperturbative results compare to perturbation
theory?  We will not attempt to do a complete one loop calculation of
the sphaleron rate in finite volume, but it is quite easy to compute the
``sphaleron'' energy, the energy of the saddle point between topological
minima.  We can use the technique of \cite{Baal}, or any other technique
which can find a saddle point solution (we have one).  We find $E =
27.77 / N$, with $N$ the linear dimension of the lattice.  In our case
that means $\beta E = 37$ and we would naively expect a rate suppression
of $\exp(-37)$ before zero modes and the fluctuation determinant are
included.  This might be compared with the free energy difference
between $N=0$ and $N=1/2$, which is of order 15.
We know that the complete inclusion of the zero modes and the
fluctuation determinant is likely to make up some of this difference,
but it certainly will not account for all of it.
Nonperturbative physics is at work and it
enhances the rate of sphaleron transitions in this case.

\subsection{broken phase rate}

Now we will apply the same technology to the physically interesting case
of large volume, broken phase SU(2)$ \times $U(1) Higgs theory.  
Since the previous subsections already explained both the Monte-Carlo
update technique and the real time tools used to find $\langle |dN/dt|
\rangle$ and the dynamical prefactor, we will just discuss here how this
case is different from the small volume, symmetric phase calculation,
and what we have to do differently to get it to work.

As we
discussed, we will work at the critical temperature, which corresponds
in the 3-D language to the critical Higgs mass.  This means that the
broken phase, which we want to study, is actually only metastable; the
symmetric phase is equally thermodynamically preferable.  In a small
volume there can be fairly easy tunneling between the two, but the
metastability becomes stronger as the 
volume becomes larger.  We must choose
a volume which is large enough that metastability is very strong and
tunneling between the phases will not occur.  This means that the
physical size of the lattice we use must be significantly larger than
the physical size of the sphaleron, which would have to be true anyway
to keep the exponential tails of the sphaleron from seeing each other
around the periodic boundary conditions.  

The problem of tunneling to the symmetric phase is made worse because
the sphaleron has a zero of the Higgs condensate at its core, so it
looks something like a symmetric phase bubble.  To keep from nucleating to 
the symmetric phase we should use a volume big enough that the free
energy of the state intermediate between phases is comparable to the
free energy of the sphaleron.  We have studied three values for $x
\equiv \lambda / g^2$, $x=0.047,$ 0.039, and 0.033; for the former two
we used a physical volume of $(16 / g^2 T)^3$ and for the latter we used
$(13.33 / g^2 T)^3$.  These were all sufficient to prevent nucleation of
the symmetric phase, but for $x=0.047$, a volume of $(12.8
/ g^2 T)^3$ was not.  The volume requirement becomes less severe as the
phase transition becomes stronger at
smaller $x$, so for $x=0.039$ and 0.033 we used a good margin of excess
volume.  The volume requirement would also have been less severe if we
worked below, rather than at, the equilibrium Higgs mass parameter.

The need for a large volume drives up numerical costs in two ways.  One
is obvious; we need to update a lot of lattice volume which is ``dead
weight'' since the sphaleron is not sitting there.  But the large volume
also makes the multicanonical algorithm perform worse.  Examining figure
\ref{compare_tau0}, we see the free energy rises roughly linearly with
$N$ at small values of $N$.  This behavior is also expected
analytically in the broken phase case, see for instance \cite{DPSSG}.
Naively, then, having the gauge fields being part way up
the sphaleron at one place in the box is not thermodynamically favored
over having them half as far up the sphaleron in two different places,
at least for relatively small $N$.  In fact the entropy from getting to
choose the locations of the two places means that this may be slightly
preferred to being further up the sphaleron in one place, for the range
of $N$ where the
free energy is varying linearly with $N$.  But at larger $N$ the free
energy for a single ``near sphaleron'' levels off while that for two
continues to rise linearly, so near 
$N=1/2$ we prefer having a single sphaleron.  Somewhere in between
there is a mismatch in what kind of configuration is dominating the
ensemble, and such a mismatch can reduce the efficiency of the
Monte-Carlo.

To cut the numerical demand we integrate out $A_0$
fields (i.e., we do not enforce Gauss' Law when we study dynamics), 
which allows us to use the very efficient update algorithm of
Rummukainen et. al. \cite{KLRSresults}.  
In fact a single step of the Rummukainen et. al. 
update (one heat bath and four overrelaxation sweeps) is far too large
an update of the fields; if the value of $f(N)$ typically changes by
more than 1 under one update then the accept rate for updates becomes
very small.  Instead, we alternate between performing a ``scaled back''
version of the update in which only some ``mod $p$ checkerboard'' of the
sites are updated, and an overrelaxation sweep updating only the 
Higgs fields.  Since the overrelaxation sweep of the Higgs fields does
not change the gauge fields, and since $N$ depends on the gauge fields
alone, this sweep can be automatically accepted.  Doing this means that
the evolution of the Higgs fields can be made ``fast'' in comparison to
the update of the gauge fields.  That ensures that the thermalization 
of the gauge fields is not gummed up by slow evolution of the Higgs 
fields.  Even with the extra updates for the Higgs fields, and the
blocking procedure for accelerating the measurement of $N$, most machine
time is spent measuring $N$, and further improvements in the update
algorithm won't help.

The results for the flux through the $N=1/2$ separatrix are given in
Table \ref{Table1}, and also in \cite{sphaleron1}.
We have also measured the dynamical prefactor for the $x=0.039$ case,
without added hard thermal loops or enforcement of Gauss' Law.  The
value we get is $.33 \pm .05$, slightly 
lower than the value for the gradient
flow separatrix, which is $0.40 \pm .05$ 
(we measure this by choosing a value of
$\tau_0$ large enough that the two separatrices almost coincide, so
there is a one to one correspondence between crossing $N=1/2$ and
crossing the gradient flow separatrix).  It is not clear 
whether this represents some interesting dynamical behavior of the
theory or whether it means that the Yang-Mills gradient flow separatrix
is not the optimal divider between topological vacua.

Another interesting question is how the dynamical prefactor depends on
hard thermal loops.  Unfortunately, the numerical cost of using the
``particles'' technique of \cite{particles} is so high in this context
that we have to cut a few more corners to make the calculation.  We drop
the U(1) factor, increase the lattice spacing from $a=2/(5g^2 T)$ to
$a=1/(2 g^2 T)$ ($\beta = 8$), and reduce the volume to $(14 / g^2
T)^3$.  To prevent nucleation to the symmetric phase, we work at a
somewhat larger value of $x$, $x=.042$, and below the equilibrium
temperature, so the broken phase is more stable than the symmetric
phase.  We choose the temperature (the thermal Higgs mass, really) 
so the Higgs condensate is $\phi_0 = 1.7 gT$, just more than enough to
make the sphaleron rate too low to erase baryon number (we expect).
We cannot directly measure the sphaleron rate with this set of
parameters because the update including particles is expensive enough
that the multicanonical calculations are prohibitive, 
but we can get a good sample of points
on the separatrix by using a reweighting which favors $N \sim 1/2$
strongly, and we can get sufficient statistics for the dynamical
prefactor to make a good determination.  

We can ``shut off'' the hard thermal loop contribution to the dynamics
without changing their contribution to thermodynamics by not allowing
the particles to move during the Hamiltonian evolutions used to
determine the dynamical prefactor.  Also, if the arguments of Arnold are
correct \cite{Arnoldlatt}, changing the velocity at which the
particles move changes their contribution to the key part of the hard
thermal loops, the $\omega \ll k \sim g^2 T$ regime, linearly in the
velocity.  Hence we could explore very strong hard thermal loops either
by putting in very many particles, or by making them move very fast.
The numerical cost is the same but the memory costs favor the latter.

We use a realistic total hard thermal loop strength, $m_D^2 = 5 g^4 T^2$
from particles.  (The standard model value is $m_D^2 = (11/6) g^2 T^2$,
and since $g^2 \sim .40$ on performing the dimensional reduction
calculation \cite{KLRS}, this is just smaller than the value we used.)
The dynamical prefactor changes only mildly, from $.52 \pm .05$ to 
$.40 \pm .05$, when we turn the particles on.  When we
increase the particle velocities to 4 times the speed of light, the
prefactor becomes $ .15 \pm .03$.  Hard thermal loops do indeed reduce
the dynamical prefactor, which is already less than 1 without them.
However, the parametric limit in which the reduction is large is not
achieved for realistic parameter values.  Also note that the value $.52$
is larger than we got at $x=.039$ and $m_{\rm H}^2 = m_{H,\rm crit}^2$.  We
assume this is because the larger value was evaluated 
below the equilibrium temperature, where
the Higgs condensate is larger and stiffer and the sphaleron is smaller
and more energetic; its decay should be more vigorous and less 
susceptible to buffeting by large IR fields.

\section{Results}
\label{Resultsec}

We present our results for the diffusion constant for $N_{\rm CS}$ in
Table \ref{Table1}.  The first quoted value is without the dynamical
prefactor.  The dynamical prefactor is less than 1 for the ($N=1/2$)
separatrix we have used, and also for the gradient flow separatrix; its
value for the $x=0.039$ data is about $0.33$, and we will take this to be
representative of the other two values of $x$ as well.  It is not clear
whether the prefactor is less than 1 because our separatrix is
sub-optimal, or because the dynamics are nontrivial in a way which often
leads to multiple crossings.

We have 
found that the dynamical prefactor depends on the strength of hard
thermal loops; but for the realistic Debye mass the effect is not
strong.  A reasonable estimate for the relevant dynamical prefactor
for the cases of interest and for the value of $\tau_0$ we used to
define the measurable $N$ is around $0.3 \pm 0.1$.  We have
included a row in the table where we include this (estimated) effect in
the rate.

\begin{figure}[t]
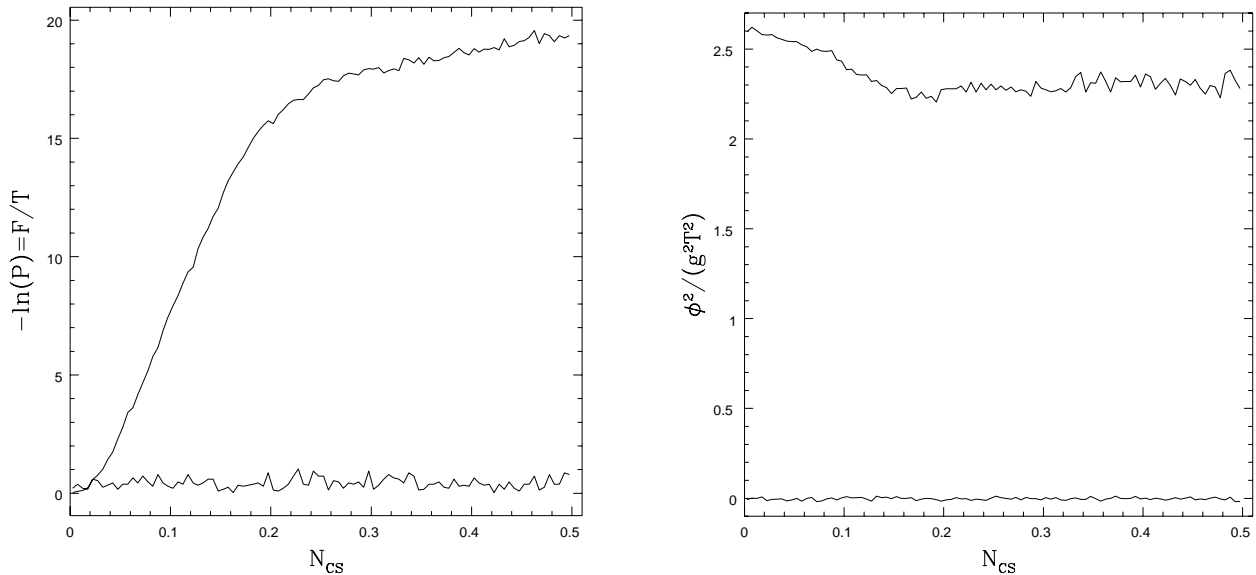

\centerline{\mbox{\psfig{file=fenergy.epsi,width=3in}} \hspace{0.4in}
\mbox{\psfig{file=phi.epsi,width=3in}}}
\caption{ \label{symm_vs_broken} 
Free energy (left) and $( \phi^\dagger \phi_{\rm
broken} - \phi^\dagger \phi_{\rm symm} ) / ( g^2 T^2 ) $ as functions of
$N$ at $x \equiv \lambda / g^2 = 0.039$, in a $(16/g^2 T)^3$
volume (at $\tau_0 = 3.6 / (g^2 T)^2$).  
In each case the upper curve is the broken phase and the lower
curve is the symmetric phase.  The plot of $\phi^2$
shows that the volume used was large enough that the
sphaleron did not bring us anywhere near a 
transition to the symmetric phase.  The behavior of the two phases is
completely different; in the broken phase there is a free energy barrier, 
and in the symmetric phase there is not.}
\end{figure}

For amusement, we have also compared the broken and symmetric phase
probability distributions for $N$.
We present the results in Figure \ref{symm_vs_broken}, which
also shows how $\phi^2$ varies with $N$ in each case.  The
two data sets were taken using identical values for all parameters
(lattice spacing $a=2/5g^2 T$, $x=0.039$, critical temperature (i.e.
critical Higgs mass), lattice volume = $ (16 / g^2 T)^3$, and $\tau_0 = 3.6 /
(g^2 T)^2$), but starting with a broken phase
initial condition in one case and a symmetric phase initial condition in
the other.  A barrier to changing $N_{\rm CS}$ is clearly present in the
broken phase case and clearly absent in the symmetric phase case.

We cannot use the data for the symmetric phase case to determine
$\Gamma_d$ in the symmetric phase.  Although it would be straightforward
to compute $\langle | dN/dt | \rangle$ and get the flux through the
separatrix, the calculation of the dynamical
prefactor is impossible.  A Hamiltonian path through the 
separatrix does not settle into the neighborhood of a topological vacuum 
and stay there for a long time; it just continues to wander around, as
we already saw in Figure \ref{jump_fig}.  But we can measure $\Gamma_d$
in the symmetric phase with purely real time techniques; the value including
hard thermal loops (and actually for pure SU(2) Yang-Mills theory) is
$\Gamma_d = (29 \pm 6) \alpha_w^5 T^4$, or $- \ln \Gamma_d \simeq
13.9$.  (This result may get revised downwards somewhat when the issues
involving logarithmic corrections \cite{Bodek_log} have been fully
accounted for.)
As expected, the broken phase rate at small $x = \lambda / g^2$ 
is enormously smaller; for $x = .033$ the ratio is about $10^8$.

\subsection{comparison to perturbation theory}

We want to compare the determined sphaleron rate to two things;
perturbation theory, and the value required to avoid erasure of baryon
number generated at the electroweak phase transition.  One loop
perturbation theory gives \cite{ArnoldMcLerran}\footnote{The 
definition of $\Gamma$ used in \protect{\cite{ArnoldMcLerran,Carson}} 
is the response rate to a chemical potential,  which is
half the diffusion rate \protect{\cite{KhlebShap}}; 
so Eq. (\protect{\ref{Gammapert}}) differs by a factor of 2
from the expressions in those references.}  
\begin{equation}
\Gamma_d = 4 T^4 \frac{\omega_{-}}{g \phi_0} \left( \frac{\alpha_W}{4
	\pi} \right)^4 \left( \frac{4 \pi \phi_0}{g T} \right)^7 
	{\cal N}_{\rm tr} {\cal NV}_{\rm rot} \kappa e^{- \beta E_{\rm
	sph}} \, .
\label{Gammapert}
\end{equation}
Here $\phi_0$ is
the broken phase Higgs condensate expectation value, $\omega_{-}$
is the unstable frequency of the sphaleron, ${\cal N}_{\rm tr} 
{\cal NV}_{\rm rot}$ are zero mode factors, $\kappa$ is the one loop
fluctuation determinant, and $E_{\rm sph}$ is the energy of the
Klinkhamer Manton sphaleron, using the tree level Hamiltonian.  
For small $\lambda / g^2 \equiv x$, $- T \ln \kappa$ equals the energy due 
to the one loop effective potential term, plus a modest correction
\cite{Baacke}.  We can guess that the dominant two loop corrections to
Eq. (\ref{Gammapert}) are absorbed by including the two loop
effective potential terms in the Hamiltonian.  So it seems reasonable to
estimate the sphaleron rate by Eq. (\ref{Gammapert}), but setting
$\kappa = 1$ and solving for the sphaleron energy using the two loop
effective potential at the equilibrium temperature.  One should also
solve for the zero modes and $\omega_{-} / \phi_0$ 
at this value, but they are very
weak functions of the effective potential \cite{Carson}.  We use the
values from \cite{Carson} at $x = (\lambda / g^2) = 0.04$ for these, but
solve for the sphaleron energy, $E_{\rm sph} = 4 \pi B \phi_0 / g$,
numerically, using the two loop effective potential at $T_{\rm c}$.  We 
use the two loop potential presented in \cite{Hebecker}, without pieces 
from longitudinal gauge bosons (assumed integrated out). We also drop
two loop terms proportional to $\lambda g^2$ or $\lambda^2$, because the
perturbative determination of $\phi_0$ is an expansion in $\lambda /
g^2$, and such terms contribute at the same or higher order as unknown 3
loop terms.  (Including these 2 loop terms moves $\phi_0$ closer to the
nonperturbative value by an insignificant amount.)  The ``two loop'' 
analytic sphaleron rate, also included in Table \ref{Table1}, 
is about $\exp(2.5)$ times faster than the numerically 
determined nonperturbative rate, and falls further off when we include
the dynamical prefactor\footnote{There is no literature calculation of
the dynamical prefactor, including hard thermal loops but in the
perturbative context.}.  The difference is more than can be 
explained by the difference in $\phi_0$, but it is not huge in the
sense that it represents a change of less than $20\%$ in the exponent.
The difference gets smaller, relative to the exponent, as the 
sphaleron energy becomes larger.

\subsection{comparison to the erasure bound}

We should compare the sphaleron rate to the limit set by
requiring that baryon number not be erased.  The rate at which
sphalerons degrade baryon number is \cite{ArnoldMcLerran}
\begin{equation}
\frac{1}{N_{\rm B}} \frac{d N_{\rm B}}{dt} = 
	- \frac{13 N_{\rm F}}{4} \Gamma_d T^{-3} \, ,
\end{equation}
where $N_{\rm F}=3$ is the number of families, and the numerical 
factor $13 N_{\rm F}/4$ would be smaller in theories, such as
supersymmetry,  in which additional degrees of freedom can store 
baryon number.\footnote{Again, there is a factor of 2 difference from 
the reference because they write in terms of the response to a chemical
potential, which is half the diffusion constant.  This 2 cancels the
other 2.}  Integrating from the end of the phase transition to the
present day,
\begin{equation}
\ln ( N_{\rm B} / N_{\rm B}(T_{\rm c}) ) 
	= - \frac{13 N_{\rm F}}{4} \int_{t_0}^\infty \Gamma_d(T(t)) 
	T^{-3}(t) dt \, ,
\end{equation}
where we have shown the dependence of $\Gamma_d$ on $T$ and of $T$ on
$t$.  

Now $\ln \Gamma_d$ is very
sensitive to $\sqrt{\langle \phi^\dagger \phi \rangle}$, and hence to
$T$; so we can
approximate $\ln \Gamma_d(T) \simeq \ln \Gamma_d(T_{\rm c}) + 
(T-T_{\rm c}) (d \ln\Gamma_d/dT)|_{T=T_{\rm c}}$, 
and perform the integral:
\begin{equation}
\ln ( - \ln ( N_{\rm B} / N_{\rm B}(T_{\rm c}) ) ) = \ln(39/4) + 
	\ln \left( \frac{\Gamma_d(T_{\rm c})}{T^{4}} \right) 
	- \ln \left( \left. - \frac{ d \ln \Gamma_d(T(t)) }{T dt} 
	\right|_{T=T_{\rm c}} \right) \, .
\label{doublelog}
\end{equation}
By the chain rule,
\begin{equation}
\frac{d \ln \Gamma_d}{T dt} = \frac{d \ln \Gamma_d}{d y} \frac{dy}{dT}
	\frac{d \ln T}{dt} \, ,
\end{equation}
where $y = m_{\rm H}^{2}(T) / (g^4 T^2)$ is the dimensionless thermal
Higgs mass squared.  

We get $dy/dT$ from the 1 loop correction to $m_{\rm H}^2$ \cite{Hebecker},
\begin{equation}
\frac{dy}{dT} \simeq \frac{8 \lambda + 4 g_{\rm y}^2 + 
	g^2 (3 + \tan^2 \Theta_W)}{8 g^4 T} \, ,
\end{equation}
and we get $d \ln T/dt$ from the Friedmann equation in a radiation
dominated universe,
\begin{equation}
\frac{1}{4 t^2} = H^2 = \frac{8 \pi G}{3} 
	\frac{ \pi^2 g_*}{30} T^4  \quad \Rightarrow \quad 
	\frac{d \ln T}{dt} = - \sqrt{ \frac{4 \pi^3 g_*}{45} }
	\frac{T^2}{m_{\rm pl}}
	\, ,
\end{equation}
where $g_*$ is the number of radiative degrees of freedom in 
the universe ($g_* = 106.75$ in the minimal standard model) 
and $m_{\rm pl} \simeq 1.22 \times 10^{19}$GeV is the Planck
mass.  Finally, we determine $d\ln \Gamma_d / dy$ perturbatively, by 
varying $y$ slightly from the equilibrium value and recomputing the 
two loop sphaleron rate.  The dependence is quite strong.  
We include it in Table \ref{Table1}.

The most widely cited discussion of baryon number erasure after the
phase transition makes the approximation that the baryon number
violation rate after the phase transition is constant for about one
Hubble time \cite{Shapwrong}.  In fact,
because $\Gamma_d$ depends very strongly on $y$, which in turn depends
strongly on $T$, most baryon number erasure occurs in the first $10^{-3}$
Hubble times after the phase transition.  Hence the initial rate of baryon
number violation,
$\Gamma_d(T_{\rm c})$, which prevents washout is $10^3$ times 
larger than assumed in \cite{Shapwrong}, leading to a weaker
bound on $\Gamma_d(T_{\rm c})$, roughly
\begin{equation}
- \ln (\Gamma_d(T_{\rm c}) T_{\rm c}^{-4} ) 
	> 30.4 - \ln ( T_{\rm c} / 100{\rm GeV} )\, .
\label{thebound}
\end{equation}
The values of $g_*$ and $dy/dT$ will both be larger in supersymmetric 
extensions of the standard model.  Unless quite a number of
supersymmetric partners have masses under 100 GeV, which now seems
unlikely, $g_*$ will not be too much larger; but if there are stops with
SUSY breaking masses of less than 100 GeV, which is necessary to get a
strong enough transition without 
violating the current experimental Higgs mass bound, then
$T dy/dT$ gets extra contributions from stops which bring it up 
by about a factor of 2.  The bound, Eq. (\ref{thebound}), is weakened
by about 1.  Also note that, because
Eq. (\ref{doublelog}) is for the double log of $N_{\rm B} / N_{\rm
B}(T_{\rm c})$, failing to meet the bound by 1 means the baryon number is
diminished by $\exp(\exp(1)) \simeq 15$, and failure by 2 reduces baryon
number by $\exp(\exp(2)) \simeq 1600$; so the bound is quite sharp.

\section{Conclusion}
\label{Conclusion}

We have shown how to define a lattice measurable which allows a
(multicanonical Monte-Carlo) nonperturbative determination of the broken
phase sphaleron barrier height.  We have combined this with real time
techniques to measure the diffusion constant of $N_{\rm CS}$ (and hence
baryon number violation) in the
broken electroweak phase nonperturbatively, including the dynamical
prefactor.  We find that the diffusion constant is smaller by
about $\exp(-3.6)$ than in a perturbative estimate using the two loop
effective potential and no wave function corrections
(and assuming a dynamical prefactor of 1).  
The difference is too large to ascribe to the difference between the 
perturbative and nonperturbative value of $\phi_0$. 
However, it represents a shift in the exponent of less than $20 \%$ from
the perturbative estimate.

We have also demonstrated that the physics of hard thermal loops does
change the sphaleron rate in the broken phase, apparently consistently
with the arguments of Arnold, Son, and Yaffe \cite{ArnoldYaffe}.  But to
really achieve the parametric limit they discuss takes an
unrealistically large Debye mass; for physical values of the parameters,
the correction due to hard thermal loops is fairly minor.

Interpolating between the values of $x \equiv \lambda / g^2$ 
where we have measured, and
including the estimate for the dynamical 
prefactor, we get a bound of about $(\lambda / g^2) \equiv x=.037$ 
in the standard model and
$x=.039$ in the MSSM when it can be perturbatively reduced to a standard
model like effective theory.  These are slightly looser than we quote in
\cite{sphaleron1} because the measured value of the dynamical prefactor
is smaller than our estimate there.  
Another convenient way to state our result is that the bound on the Higgs
condensate after the phase transition is about $\phi_0 = 1.67 gT$ in the
MSM and $\phi_0 = 1.60 gT$ in the MSSM (with the MSSM value weaker 
because the larger temperature dependence of the thermal Higgs mass
makes the erasure rate fall off faster with time after the transition).

The bound is softened if the universe does not reheat to $T_c$ during 
the phase transition, because the phase transition then ends at a lower
temperature with a larger Higgs vev.  Incomplete reheating 
may well be generic.  Unfortunately, we do
not have a nonperturbative measurement of the bubble nucleation action,
which is required to determine definitively whether reheating happens.
This is an interesting project which can perhaps be approached by
techniques similar to what we have used here, i.e. defining a separatrix
corresponding to the critical bubble (as the separatrix here corresponds
to the sphaleron), and making a complete calculation including the
dynamical prefactor.

We should comment on what we expect to be true below the
equilibrium temperature and in the case of a light stop.  As the
temperature falls below equilibrium and the Higgs condensate becomes
larger, perturbation theory should become better, at least in the sense
that the error in the exponent should get smaller compared to the
magnitude of the exponent.  It would be very surprising if the
nonperturbative rate switches to being faster than the perturbative
estimate, though.  It is straightforward, though expensive, to repeat
the analysis here for temperatures below the equilibrium point.
Probably it will only be worth it after we have nonperturbative
information on the bubble nucleation rate.  In the case of the light
stop, we expect the most important differences to be in the effective
potential, in which case perturbation theory should do as well as it
does here, if we permit a ``by hand'' correction for the strength of the
phase transition.  We expect this because, while the stop contributes
at one loop to the effective potential, scalar interactions only influence
wave function corrections (say, the Higgs gradient energy of the
sphaleron) at two loops, and there is no direct interaction between a
light right stop and the SU(2) gauge fields.

\section*{Acknowledgments}

I would like to thank Jim Cline, Jim Hetrick, Kari Rummukainen,
Misha Shaposhnikov, and Alex Krasnitz for 
discussions and correspondence.  I also thank 
Berndt M\"{u}ller and the North Carolina Supercomputing Center.

\end{document}